\documentclass[aps,preprint,superscriptaddress,nofootinbib,eqsecnum]{revtex4-1}
\usepackage{amsmath,amssymb}
\usepackage{graphicx}
\usepackage{epsfig}
\usepackage{graphicx}
\usepackage{epstopdf}
\usepackage{amsfonts}
\usepackage{amsmath}
\usepackage{amssymb}
\usepackage{enumerate}%
\usepackage{subfigure}
\usepackage{color}
\newcommand{\be}{\begin{equation}}
\newcommand{\ee}{\end{equation}}
\newcommand{\bea}{\begin{eqnarray}}
\newcommand{\eea}{\end{eqnarray}}
\newcommand{\ba}{\begin{array}}
\newcommand{\ea}{\end{array}}

\newcommand{\eq}[1]{(\ref{#1})}
\newcommand{\bes}{\begin{subequations}}
\newcommand{\ees}{\end{subequations}}

\newcommand{\Tr}{\mbox{Tr}}

\begin{document}

\preprint{}

\title{Testing Information Causality for General Quantum Communication Protocols}

\vfill

\author{ I-Ching Yu\footnote{896410029@ntnu.edu.tw}, Feng-Li Lin\footnote{linfengli@phy.ntnu.edu.tw, the corresponding author}}
\affiliation{Department of Physics, National Taiwan Normal
University, Taipei, 116, Taiwan}

\vfill

\begin{abstract}

   Information causality was proposed as a physical principle to put upper bound on the accessible information gain in a physical bi-partite communication scheme. Intuitively, the information gain cannot be larger than the amount of classical communication to avoid violation of causality.
   Moreover, it was shown that this bound is consistent with the Tsirelson bound for the binary quantum systems.
     In this paper, we test the information causality for the more general (non-binary) quantum communication schemes. In order to apply the semi-definite programming method to find the maximal information gain, we only consider the schemes in which the information gain is monotonically related to the Bell-type functions, i.e., the generalization of CHSH functions for Bell inequalities in a binary schemes. We determine these Bell-type functions by using the signal decay theorem.  Our results support the proposal of information causality.  We also find the maximal information gain by numerical brute-force method for the most general 2-level and 2-setting quantum communication schemes. Our results show that boundary for the information causality bound does not agree with the one for the Tsirelson bound.
\end{abstract}


\maketitle

\tableofcontents

\*\\
\section{Introduction}\label{intro}

   The advantage of quantum information has been well exploited in improving the efficiency and reliability for the computation and communication  in the past decades. However, even with the help of the seemingly non-local quantum correlation resources,  the trivial communication complexity still cannot be reached. The communication complexity could be understood as the bound on the  accessible information gain between sender and receiver.  Recently, this bound on the information gain is formulated as a physical principle, called the information causality. It states that the information gain in a {\it physical} bi-partite communication scheme cannot  exceed the amount of classical communication.  Intuitively, this is a reasonable and physical constraint. Otherwise, one can predict what your distant partite tries to hide from you and do something to violate causality. For some particular   communication schemes with physical resources shared between sender and receiver, it was shown  \cite{IC,Our} that  the bound from the information causality is equivalent to the Tsirelson bound \cite{T2} for the binary quantum systems.

     By treating information causality as a physical principle, one can disqualify some of the no-signaling theories \cite{no-signal} from being the physical theories if they yield the results violating the information causality. In this way, it may help to single out quantum mechanics as a physical theory by testing the information causality for all possible quantum communication schemes. For example, some efforts along this line was done in \cite{dIC}.

         However, most of the tests on the information causality were performed only for the binary communication schemes. It is then interesting to test the information causality for the more general communication schemes. In this paper we will perform the testes for the d-level\footnote{The d-level here means a digit with $d$ possible values. For $d=2$ it is the usual binary digit.} quantum systems, with the more general communication protocols and the more general physical resources shared between sender and receiver. Our results agree with the bound set by the information causality. In the rest of Introduction, we will briefly review the concept of information causality to motivate this work and also outline the strategy of our approach.

     Information causality can be presented through the following task of random access code (RAC): Alice has a database of $k$ elements, denoted by the vector $\vec{a}= (a_{0},a_{1},,,a_{k-1})$.
Each element $a_{i}$ is a d-level digit (dit) and is only known to Alice. A second distant party, Bob is given a random variable $b\in{0,1,2,,,k-1}$. The value of $b$ is used to instruct Bob in
guessing the dit $a_b$ optimally after receiving a dit $\alpha$ sent by Alice. In this context, the information causality can be formulated as follows:
\begin{equation}
I=\sum_{i=0}^{k-1}I(a_{i};\beta |b=i)\leq \log_{2}d\;.  \label{ic-1}
\end{equation}
where $I(a_{i};\beta |b=i)$ is Shannon's mutual information between $a_i$ and Bob's guessing dit $\beta$ under the condition $b=i$. Then, $I$ is the information gain of the communication scheme which is bounded by the amount of the classical communication encoded in $\alpha$.

\begin{figure}[t]
\includegraphics[width=0.75\columnwidth]{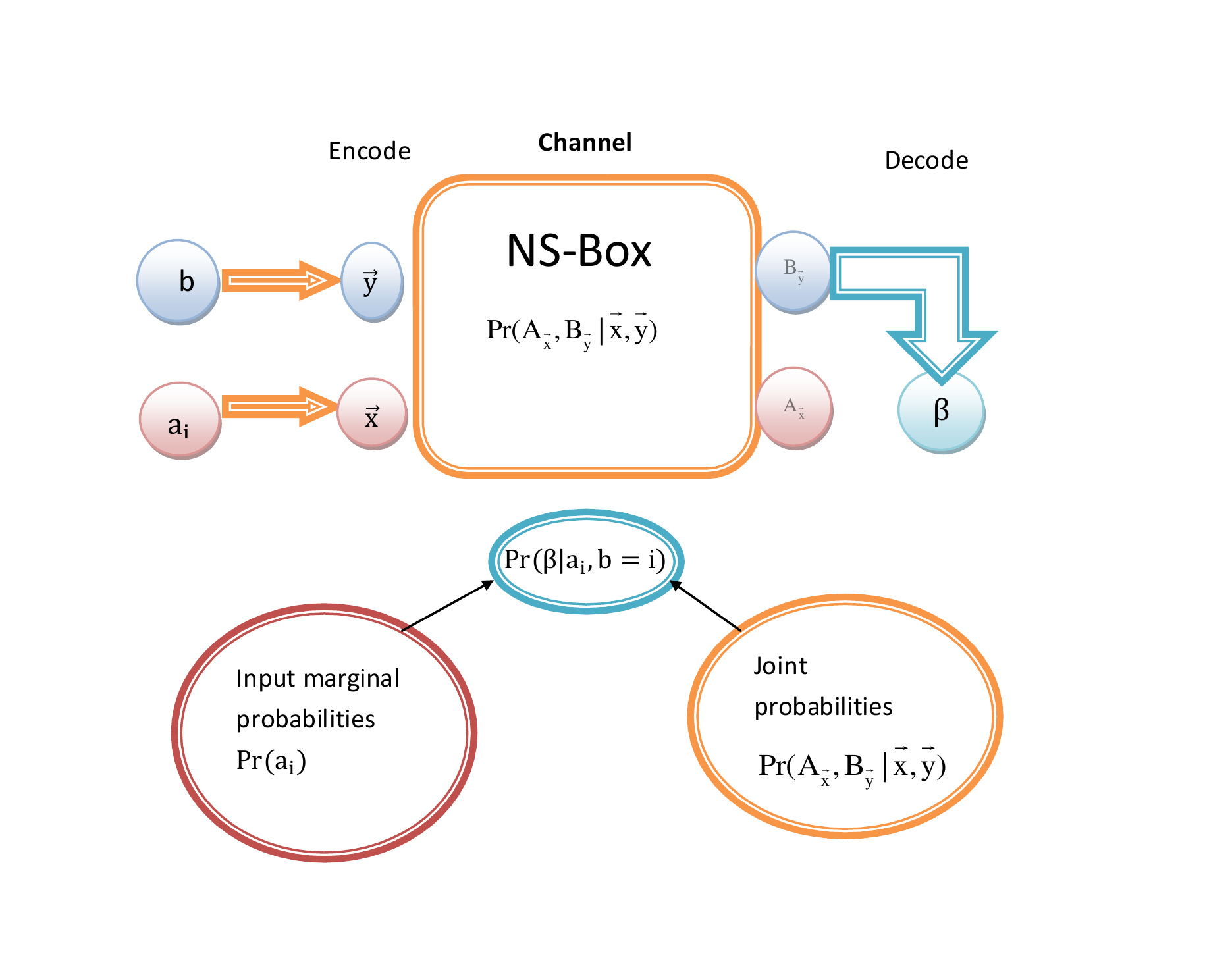}\newline
\caption{Ingredients of the communication schemes considered in this paper} \label{NS-box}
\end{figure}

    The above  information gain $I$ is determined by three parts of the communication scheme: (1) the exact RAC protocol, (2) the communication channel and (3) the input marginal probabilities denoted by  $\Pr(a_{i})$.  This is shown in Fig \ref{NS-box}.  The purpose of RAC encoding is for Alice to encode her data $\vec{a}$ into $\vec{x}$ and Bob to do his $b$ into $\vec{y}$. The details will be given in section \ref{sec2}.

     The second part in our communication scheme is a given channel specified by the pre-shared correlation between Alice and Bob, the so-called no-signaling box (NS-box).  The aforementioned encoded data $\vec{x}$ and $\vec{y}$ are the input of the NS-box which then yields the corresponding outputs $A_{\vec{x}}$ and $B_{\vec{y}}$, respectively.   Bob will then combine $B_{\vec{y}}$ with  the classical information sent from Alice to guess  $\vec{a}$.  Most importantly, the NS-box is characterized by the conditional joint probabilities $\Pr(A_{\vec{x}},B_{\vec{y}}|\vec{x},\vec{y})$, and should satisfy the following no-signaling condition \cite{no-signal}
\begin{equation}
\sum_{B_{\vec{y}}}
\Pr(A_{\vec{x}},B_{\vec{y}}|\vec{x},\vec{y})=\Pr(A_{\vec{x}}|\vec{x})\quad
\mbox{and}\quad
\sum_{A_{\vec{x}}}\Pr(A_{\vec{x}},B_{\vec{y}}|\vec{x},\vec{y})=\Pr(B_{\vec{y}}|\vec{y}),\qquad
\forall \vec{x},\vec{y}. \label{ns-box}
\end{equation}
 This implies that superluminal signaling is impossible.

    Now comes the third part in our communication scheme: the input marginal probabilities. They are usually assumed to be uniform and not treated as variables. However, when evaluating information gain  $I$ in \eq{ic-1}, we need the conditional probabilities  $\Pr(\beta|a_i,b=i)$, which are related to both the joint probabilities $\Pr(A_{\vec{x}},B_{\vec{y}}|\vec{x},\vec{y})$ of the NS-box and the input marginal probabilities $\Pr(a_i)$.  In this work, we will consider the more general communication schemes with variable and non-uniform $\Pr(a_i)$ and evaluate the corresponding information gain.

      Naively, one would like to find the information gain of our communication schemes by maximizing the information gain $I$ over  $\Pr(a_i)$ and $\Pr(A_{\vec{x}},B_{\vec{y}}|\vec{x},\vec{y})$. The joint probabilities of the NS-box $\Pr(A_{\vec{x}},B_{\vec{y}}|\vec{x},\vec{y})$ should be realized by the quantum correlations. However, we will show that this maximization problem is not a convex problem so that it cannot be solved by numerical recipes.

   To by-pass this no-go situation, we choose two ways to proceed. The first
 way is to consider an alternative convex optimization problem, whose object function and the  information gain $I$ are monotonically related under some special assumptions. It turns out that the  alternative convex optimization problem is to find the maximal quantum violation of the Bell-type inequality. This can be thought as finding the generalized Tsirelson bound. We will call the corresponding inequality for the generalized Tsirelson bound \footnote{Note the original Tsirelson bound is only for binary quantum system. Here we consider the general cases.} the Tsirelson-type inequality, or simply the Tsirelson inequality. Correspondingly, the object function is the LHS of the Bell-type inequality, which we will call the Bell-type function, or simply Bell function.

  For the binary 2-setting communication schemes, the Bell-type function is the famous CHSH function. However, for the general schemes one should try to find the appropriate Bell-type functions. In this paper, we generalize the construction method developed in \cite{Our} to obtain such Bell-type functions. This method is based on the signal decay theorem proposed in \cite{Evans1, Evans2}. We further show that these Bell-type functions are monotonically related to $I$ for the communication schemes with unbiased (i.e., symmetric and isotropic) $\Pr(\beta|a_{i},b=i)$ and  i.i.d. inputs $\{a_i\}$ with uniform $\Pr(a_i)$.  Therefore, for such schemes we can  optimize the information gain $I$ by applying the semi-definite programing (SDP) method   \cite{Acin,Acinl} to obtain the maximum of the Bell-type function for the quantum communication schemes, i.e., the Tsirelson bound.

   On the other hand, if we would like to consider the more general communication schemes rather than the aforementioned ones so that the above monotonic relation between $I$ and the object function fails, then we will use the second way.
  This is just to maximize the information gain $I$ over $\Pr(a_i)$ and $\Pr(A_{\vec{x}},B_{\vec{y}}|\vec{x},\vec{y})$ by brutal force numerically without relying on the convex optimization.
   As limited by the power of our computation facilities, we will only consider the binary 2-setting communication schemes. Our results show that the bound required by the information causality is not saturated  by the scheme saturating the Tsirelson bound. Instead, it is saturated by the case saturating the CHSH inequality.

   The paper is organized as follows. In the next section we will define our communication schemes in details and then derive the Bell-type functions for the schemes with unbiased $\Pr(\beta|a_{i},b=i)$ and i.i.d. inputs with  uniform  $\Pr(a_{i})$. In section \ref{sec3}, we will show that maximizing the information gain $I$ over $\Pr(A_{\vec{x}},B_{\vec{y}}|\vec{x},\vec{y})$ and $\Pr(a_{i})$ is not a convex optimization problem. We also prove that the Bell-type functions and the information gain $I$ are monotonically related under some assumptions.
    In section \ref{sec4}, we briefly review the semidefinite programming (SDP) proposed in \cite{Acin,Acinl}, and then apply it to solve the convex optimization problem and find out the generalized Tsirelson bound. We use the result to evaluate the corresponding information gain $I$ and compare with the bound required by  the information causality.
     In \ref{sec5}, we will use the numerical brute-force method to maximize $I$ for general binary 2-setting schemes. Finally, we conclude our paper in section \ref{sec6}  with some discussions. Besides, several technical detailed results are given in the Appendices.

\section{The generalized Bell-type functions from the signal decay theorem}\label{sec2}

    In the Introduction, we have briefly described our communication scheme. Here we describe the details of the encoding/decoding in the RAC protocol: Alice encodes her data $\vec{a}$ as $\vec{x}:=(x_1,\cdots,x_{k-1})$ with $x_i=a_i-a_0$, and Bob does his input $b$ as $\vec{y}:=(y_1,\cdots,y_{k-1})$ with $y_i=\delta_{b,i}$ for $b\ne 0$ and $\vec{y}=0$ for $b=0$.
The dit-string $\vec{x}$ and $\vec{y}$ are the inputs of the NS-box.
The corresponding outputs of the NS-box are $A_{\vec{x}}$ and
$B_{\vec{y}}$, respectively. More specifically, the dit sent by
Alice is $\alpha=A_{\vec{x}}-a_0$, and the pre-shared correlation is
defined by the conditional probabilities
$\Pr(B_{\vec{y}}-A_{\vec{x}}=\vec{x}\cdot \vec{y}|\vec{x},\vec{y})$
between the inputs and outputs of the NS-box. Accordingly, Bob's
optimal guessing dit $\beta$ can be chosen as $B_{\vec{y}}-\alpha$.
This is because $\beta=B_{\vec{y}}-A_{\vec{x}}+a_0=\vec{x}\cdot
\vec{y}+a_0$ as long as $B_{\vec{y}}-A_{\vec{x}}=\vec{x}\cdot
\vec{y}$ holds. In this case, Bob guesses $a_b$ perfectly. Take
$d=3$ and $k=3$ as an example for illustration: Bob's optimal guess
bit is \be \beta=\vec{x}\cdot\vec{y}
+a_{0}=(a_{1}-a_{0},a_{2}-a_{0})\cdot (y_{0},y_{1})+a_{0}. \ee
 If Bob's input $\vec{y}=(y_{0},y_{1})=(0,0)$,
$\beta=a_{0}$; if $\vec{y}=(y_{0},y_{1})=(1,0)$, $\beta=a_{1}$; and
if $\vec{y}=(y_{0},y_{1})=(0,1)$, $\beta=a_{2}$. Bob can guess
$a_{b}$ perfectly.

Using the above RAC protocol, Alice and Bob have $d^{k-1}$ and $k$
measurement settings, respectively. Each of the measurement
settings will give $d$ kinds of outputs. However, the noise of the
NS-box affects the successful probability so that Bob can not always
guess $a_{b}$ correctly. If the NS-box is a quantum mechanical one,
then the conditional probabilities
$\Pr(B_{\vec{y}}-A_{\vec{x}}=\vec{x}\cdot \vec{y}|\vec{x},\vec{y}])$
should be constrained by the
Tsirelson-type inequalities, so are the joint probabilities
$\Pr(A_{\vec{x}},B_{\vec{y}}|\vec{x},\vec{y})$. Then the question is
how? For $d=2$ and $k=2$, the quantum constraint comes from the
well-known Tsirelson inequality. That is, the maximal quantum
violation of the CHSH inequality is $2\sqrt{2}$, i.e.,
$|C_{0,0}+C_{0,1}+C_{1,0}-C_{1,1}|\leq2\sqrt{2}$. Note that, each
term of CHSH function $C_{\vec{x},\vec{y}}$ can be expressed in
terms of joint probabilities as
$\Pr(00|\vec{x},\vec{y})-\Pr(01|\vec{x},\vec{y})-\Pr(10|\vec{x},\vec{y})+\Pr(11|\vec{x},\vec{y})$.
Therefore, this is the constraint for
$\Pr(A_{\vec{x}},B_{\vec{y}}|\vec{x},\vec{y})$ to be consistent with
quantum mechanics.

   However, there is no known  Tsirelson-type inequalities for the cases with $d>2$. Despite that, in \cite{Our}, we find a systematic way to construct $d=2$ and $k\geq2$ Tsirelson-type inequalities by the signal decay theorem \cite{Evans1,Evans2}.
    We will generalize this method to $d>3$ case to yield suitable Bell-type functions.
    To proceed, we first recapitulate the derivation for $d=2$ cases.

Signal decay theory quantifies the loss of mutual information when
processing the data through a noisy channel. Consider a cascade of
two communication channels: $X \hookrightarrow Y \hookrightarrow Z$,
then intuitively  we have \be I(X;Z)\leq I(X;Y). \ee Moreover,  if
the second channel is a binary symmetric one, i.e.,
\begin{equation}
\Pr(Z|Y)=\left(
\begin{array}{cc}
\frac{1}{2}(1+\xi) & \frac{1}{2}(1-\xi) \\
\frac{1}{2}(1-\xi) & \frac{1}{2}(1+\xi)
\end{array}
\right),  \notag  \label{ek}
\end{equation}
then the signal decay theorem says \be\label{signaldecayth}
\frac{I(X;Z)}{I(X;Y)}\le \xi^2. \ee This theorem has been proven to yield
a tight bound in \cite{Evans1,Evans2}.  Note that the equality is held
only when $\Pr(Y|X=0)$ and $\Pr(Y|X=1)$ are almost
indistinguishable. For more detail, please see appendix A.

In \cite{Our}, we set $X=a_{i}$, $Y=a_{0}+\vec{x}\cdot\vec{y}$
and $Z=\beta$. By construction, the bit $a_{i}$ is encoded as
$a_{0}+\vec{x}\cdot\vec{y}$ such that
$I(a_{i};a_{0}+\vec{x}\cdot\vec{y})=1$.
 Using the tight bound of \eq{signaldecayth},  we can get
\begin{equation}
I(a_{i}\;;\beta |b=i)\le  \xi _i^{2}. \label{ub}
\end{equation}
For our RAC protocol, the index of the $\xi_i$ is the vector $\vec{y}$. It is then easy to see that $\xi_{\vec{y}}$ is related to both the input marginal
probabilities $\Pr(a_{i})$ and the joint probabilities of the NS-box by
\begin{equation}
\frac{1+\xi_{\vec{y}}}{2}=\sum_{\{\vec{x}\}}\Pr(\vec{x})\Pr
\left(B_{\vec{y}}- A_{\vec{x}}=\vec{x}\cdot
\vec{y}|\vec{x},\vec{y}\;\right).
\end{equation}

Assuming that Alice's database is i.i.d., we can then sum over
all the mutual information between $\beta$ and $a_{i}$ to arrive
\begin{equation}
\sum_{i}I(a_{i}\;;\beta |b=i)\le \sum_{i} \xi _i^{2}.
\end{equation}
Though the object on the RHS is quadratic,
 we can linearize it by the Cauchy-Schwarz inequality, i.e., $| \sum_i \xi_i | \le \sqrt{k \sum_i \xi_i^2}$.
   For $d=k=2$ case with uniform input marginal probabilities $\Pr(a_{i})$, it is easy to show that $\sum_i \xi_i \le \sqrt{2}$ (or $\sum_i \xi_i^2 \le 1$) is nothing but the conventional Tsirelson inequality.
    Moreover, in \cite{Our} we use the SDP algorithm in \cite{Wehner} to generalize to $d=2$ and $k>2$ cases and show that the corresponding Tsirelson-type inequality
    is
\begin{equation}\label{lixi}
\sum_{i} \xi _i\leq\sqrt{k}.
\end{equation}
This is equivalent to say $\sum_i \xi^2 \le 1$. From the signal
decay theorem  \eq{ub} this implies that the maximal
information gain in our RAC protocol with the pre-shared quantum resource
is consistent with the information causality \eq{ic-1}.

  We now generalize the above construction to $d>2$ cases.
  First, we start with $d=3$ case by considering a cascade of two channels $X \hookrightarrow Y \hookrightarrow Z$ with the second one a 3-input, 3-output symmetric channel.
 Again, we want to find the upper bound of $\frac{I(X;Z)}{I(X;Y)}$.
 In the Appendix \ref{app-1} we show that the ratio reaches an upper bound whenever three conditional probabilities $\Pr(Y|X=i)$ with $i=0,1,2$ are almost
 indistinguishable.
   Moreover, it can be also shown that the upper bound of the ratio is again given by \eq{signaldecayth} for the symmetric channel between $Y$ and $Z$ specified by
\begin{eqnarray}
\Pr(Z|Y)=
      \left ( \begin{array}{ccc}
\frac{2\xi+1}{3}& \frac{1-\xi}{3}& \frac{1-\xi}{3}\\
\frac{1-\xi}{3} & \frac{2\xi+1}{3}&\frac{1-\xi}{3}\\
\frac{1-\xi}{3} & \frac{1-\xi}{3}&\frac{2\xi+1}{3}
\end{array}  \right).
\end{eqnarray}

  One can generalize the above to the higher $d$ cases for the symmetric channel between $Y$ and $Z$ specified as follows:
  $\Pr(Z=i|Y=i)=\frac{(d-1)\xi+1}{d}$
and $\Pr(Z=s\ne i|Y=i)=\frac{1-\xi}{d}$ with $i\in\{0,1,...,d-1\}$.
 Again we will arrive \eq{signaldecayth}.
 Based on the signal decay theorem with  $X:=a_{i}$,
$Y:=a_{0}+\vec{x}\cdot\vec{y}$ and $Z:=\beta$ and assuming that
Alice's input probabilities are i.i.d., we can sum over all the
mutual information between each $a_{i}$ and $\beta$ and obtain
\begin{eqnarray}\label{dxi}
\sum_{i=0}^{k-1}I(\beta;a_{i}|b=i)\leq\sum_{i=0}^{k-1}\xi_{i}^{2}\log_{2}(d).
\end{eqnarray}

In our RAC protocol, the noise parameter
$\xi_{\vec{y}}$ (or $\xi_i$) can be expressed as
\begin{equation}\label{xi}
\xi _{\vec{y}}=\frac{d \sum_{\vec{x}}\Pr(\vec{x}) \Pr(B_{\vec{y}}-
A_{\vec{x}}=\vec{x}\cdot \vec{y}|\vec{x},\vec{y}) -1}{d-1}.
\end{equation}

  As for the $d=2$ case, we assume the upper bound of \eq{dxi} is capped by the information causality to yield a quadratic constraint on the noise parameters.  Again, using the Cauchy-Schwarz inequality to linearize the quadratic constraint, we find
$\sum_{\vec{y}} \xi_{\vec{y}}\leq \sqrt{k}$.  Especially, if the input marginal probabilities $\Pr(a_{i})$ are uniform, then this inequality  yields a constraint on
$\Pr(A_{\vec{x}},B_{\vec{y}}|\vec{x},\vec{y})$.   Using \eq{xi}, the LHS of this inequality can be thought as a Bell-type function, and our task is to check if the RHS matches with the Tsirelson bound or not.

Then, it is ready to ask the question: If the joint probabilities of a NS-box achieve the Tsirelson bound, does the same NS-box used in our RAC protocol also saturate the information causality bound? Next, we are going to address this question.

\section{Convexity and  information gain}\label{sec3}
 \subsection{Feasibility for maximizing information gain by convex optimization}

In order to test the information causality for more general communication schemes, we have to
maximize the information gain $I$ over the conditional probabilities
$\Pr(\beta|a_{i},b=i)$ determined by the joint probabilities
$\Pr(A_{\vec{x}},B_{\vec{y}}|\vec{x},\vec{y})$ and $\Pr(a_{i})$. One
way to achieve this task is to formulate the problem as a convex optimization
programming, so that we may exploit some numerical recipes such as
\cite{CVXOPT} to carry out the task.

Minimizing a function with the equality or inequality constraints is
called convex optimization. The object function could be linear or
non-linear. For example, SDP is a kind of convex optimization with a linear object function.
Regardless of linear or non-linear object functions, the minimization (maximization) problem requires them to be convex (concave).
    Thus, if we define the  information gain $I$ as the object function for maximization in the context of information causality,  we have to check if it is concave.

 A concave function $f(x)$ ($f: \mathbb{R}^n
\rightarrow \mathbb{R}$) should satisfy the following condition:
\begin{equation}\label{convex}
f(\lambda x_{1}+(1-\lambda) x_{2})\geq \lambda
f(x_{1})+(1-\lambda)f(x_{2}),
\end{equation}
where $x_{1}$ and $x_{2}$ are $n$-dimensional real vectors, and
$0<\lambda<1$.

 Mutual information between input $X$ and output $Z$ can be written as
\begin{equation}\label{I}
I(X;Z)=H(Z)-H(Z|X)\\
      =H(Z)-\sum_{i} \Pr(X=i)H(Z|X=i),
\end{equation}
where $H(Z)=-\sum_{i}\Pr(Z=i)\log_{2}\Pr(Z=i)$ is the entropy
function. We will study the convexity of $I(X;Z)$  by varying over
the marginal probabilities $\Pr(X)$ and the channel probabilities
$\Pr(Z|X)$.

The following theorem is mentioned in \cite{Tomas}. If we fix the
channel probabilities $\Pr(Z|X)$ in \eq{I}, then $I(X;Z)$ is a
concave function with respect to $\Pr(X)$. This is the usual way in
obtaining the channel capacity, i.e., maximizing information gain
$I$ over the input marginal probabilities for a fixed channel.

   However, in the context of information causality,  the conditional probabilities $\Pr(\beta|a_{i},b=i)$ (or $\Pr(Z|X)$) are related to both the joint probabilities of the NS-box and the input marginal probabilities $\Pr(a_{i})$. This means that the above twos will be correlated if we fix $\Pr(\beta|a_{i},b=i)$.
 This cannot fit to our setup in which we aim to maximize the information gain $I$ by varying over the joint probabilities of NS-box and the input marginal probabilities $\Pr(a_{i})$.
  For example, in $d=2$ and $k=2$ case, $\Pr(\beta|a_{i},b=i)$ is given by
\begin{equation}
\Pr(\beta|a_{i},b=i)=\left(
\begin{array}{cc}
\alpha_{i} & 1-\alpha_{i} \\
1-\lambda_{i} & \lambda_{i}
\end{array}
\right) .  \notag  \label{d2c}
\end{equation}
where
\begin{eqnarray}\label{d200channel}
\alpha_{0}:=\Pr(\beta=0|a_0=0,b=0)&&=\sum_{\ell=0}^{1}\Pr(B_y-A_x=0|x=\ell,y=0)\Pr(a_{1}=\ell),
\\
\lambda_{0}:=\Pr(\beta=1|a_{0}=1,b=0)&&=\sum_{\ell=0}^1\Pr(B_y-A_x=0|x=\ell,y=0)\Pr(a_{1}=1-\ell),\\
\alpha_{1}:=\Pr(\beta=0|a_{1}=0,b=1)&&=\sum_{\ell=0}^1\Pr(B_y-A_x=\ell|x=\ell,y=1)\Pr(a_{0}=\ell),\\
\lambda_{1}:=\Pr(\beta=1|a_{1}=1,b=1)&&=\sum_{\ell=0}^1\Pr(B_y-A_x=\ell|x=\ell,y=1)\Pr(a_{0}=1-\ell).
\end{eqnarray}
From the above, we see that $\Pr(\beta|a_i,b=i)$ cannot be fixed by
varying over $\Pr(B_y-A_x|x,y)$ and $\Pr(a_{i})$ independently.
Similarly, for higher $d$ and $k$ protocols, we will also have the
constraints between the above three probabilities. Thus, maximizing
the information gain for the information causality is different from the usual
way of finding the channel capacity.

   To achieve the goal of maximizing the information gain $I$ over the input marginal probabilities $\Pr(a_{i})$ and the joint probabilities $\Pr(B_{\vec{y}}-A_{\vec{x}}|\vec{x},\vec{y})$ which can be realized by quantum mechanics, we should check if it is a convex (or concave) optimization
problem or not. If yes, then we can adopt the numerical recipe
as \cite{CVXOPT} to carry out the task. Otherwise, we can either
impose more constraints for our problem or just do it by brutal
force. It is known that \cite{Hassian} one can check if maximizing
function $f(y_1,\cdots,y_n)$ over $y_i$'s is a concave problem or
not by examining its Hessian matrix
\begin{align}
H(f)=\left(
       \begin{array}{cccc}
         \frac{\partial^{2} f}{\partial y_{1}^{2}} & \frac{\partial^{2} f}{\partial y_{1}y_{2}} & \cdots & \frac{\partial^{2} f}{\partial y_{1}y_{n}} \\
         \frac{\partial^{2} f}{\partial y_{2}y_{1}} & \frac{\partial^{2} f}{\partial y_{2}^{2}} & \cdots & \frac{\partial^{2} f}{\partial y_{2}y_{n}} \\
         \vdots & \vdots & \ddots & \vdots \\
         \frac{\partial^{2} f}{\partial y_{n}y_{1}} & \frac{\partial^{2} f}{\partial y_{n}y_{2}} & \cdots & \frac{\partial^{2} f}{\partial y_{n}^{2}} \\
       \end{array}
     \right).
\end{align}
For the maximization to be a concave problem, the Hessian matrix
should be negative semidefinite. That is, all the odd order
principal minors of $H(f)$ should be negative and all the even order
ones should be positive. Note that each first-order principal minor
of $H(f)$ is just the second derivative of $f$, i.e. ${\partial^2 f
\over \partial y_i^2}$.  So, the problem cannot be concave if
${\partial^2 f \over \partial y_i^2}>0$ for some $i$.

   With the above criterion, we can now show that the problem of maximizing $I$ over
$\Pr(B_{\vec{y}}-A_{\vec{x}}|\vec{x},\vec{y})$ and $\Pr(a_{i})$
cannot be a concave problem. To do this, we rewrite the
information gain $I$ defined in \eq{ic-1} as following:
\begin{equation}
I
=\sum_{i=0}^{k-1}\sum_{n=0}^{d-1}\sum_{j=0}^{d-1}\Pr(\beta=n,a_{i}=j|b=i)\log_{2}\frac{\Pr(\beta=n,a_{i}=j|b=i)}{\Pr(\beta=n|b=i)\Pr(a_i=j)}.\label{ijoint}
\end{equation}
Furthermore, one can express the above in terms of
$\Pr(B_{\vec{y}}-A_{\vec{x}}|\vec{x},\vec{y})$ and $\Pr(a_{i})$ by
the following relations
\begin{align}\label{pjoint}
\Pr(\beta=n, a_i=j|b=i)&=\sum_{\{a_{k\ne i}\}} \Pr(B_{\vec{y}}-A_{\vec{x}}=n-a_0|\vec{x},\vec{y}) \Pr(a_i=j) \; \Pi_{k\ne i} \Pr(a_k), \\
\Pr(\beta=n|b=i)&=\sum_{j=0}^{d-1}\Pr(\beta=n, a_{i}=j|b=i),
\end{align}
where $\vec{x}$ and $\vec{y}$ in the above are given by the RAC encoding, i.e.,  $\vec{x}:=(x_1,\cdots,x_{k-1})$ with
$x_i=a_i-a_0$ and $\vec{y}:=(y_1,\cdots,y_{k-1})$ with
$y_i=\delta_{b,i}$ for $b\ne 0$ and $\vec{y}=0$ for $b=0$.

    Moreover, both  $\Pr(B_{\vec{y}}-A_{\vec{x}}|\vec{x},\vec{y})$ and $\Pr(a_i)$ are subjected to the normalization conditions of total probability. Thus we need to solve these conditions such that the information gain $I$ is expressed as the function of independent probabilities. After that, we can evaluate the corresponding Hessian matrix to examine if the maximization of $I$ over these probabilities is a concave problem or not.

 For illustration, we first consider the $d=2$ and $k=2$ case. By using the relations \eq{pjoint} and the normalization conditions of total probability to implement the chain-rule while taking derivative,  we arrive
\begin{eqnarray}\label{d2I}
&& \frac{\ln 2 \cdot \partial^{2}I}{\partial(\Pr(B_y-A_x=0|x=0,y=0))^2}=\notag\\
&&-(\frac{1}{\Pr(\beta=0|b=0)}+\frac{1}{\Pr(\beta=1|b=0)})(\Pr(a_{0}=0)\Pr(a_{1}=0)-\Pr(a_{0}=1)\Pr(a_{1}=1))^{2}\notag\\
&&+(\Pr(a_{0}=0)\Pr(a_{1}=0))^{2}(\frac{1}{\Pr(\beta=0, a_{0}=0|b=0)}+\frac{1}{\Pr(\beta=1, a_{0}=0|b=0)})\notag\\
&&+(\Pr(a_{0}=1)\Pr(a_{1}=1))^{2}(\frac{1}{\Pr(\beta=0,
a_{0}=1|b=0)}+\frac{1}{\Pr(\beta=1, a_{0}=1|b=0)}).
\end{eqnarray}
Obviously, \eq{d2I} cannot always be negative. This can be seen
easily if we set $\Pr(a_0)=1-\Pr(a_1)$ so that the first term on the
RHS of \eq{d2I} is zero. Then, the remaining terms are non-negative
definiteness. This then indicates that maximizing $I$ over the joint
probabilities is not a concave problem.

    The check for the higher $d$ and $k$ cases can be done similarly, and the details can be found in the Appendix B. Again, we can set all the $\Pr(a_i)$ to be uniform so that we have
\begin{eqnarray}\label{dI3}
&&\frac{ d^{2k}\ln2 \cdot \partial^{2} I}{\partial(\Pr(B_{\vec{y}}-A_{\vec{x}}=0|\vec{x}=\vec{0},\vec{y}=\vec{0}))^2}=\notag\\
&&\sum_{n=0}^{d-1}(\frac{1}{\Pr(a_{0}=n,\beta=n|b=0)}+\frac{1}{\Pr(a_{0}=n,\beta=n+1-d|b=0)})>
0.
\end{eqnarray}

 \subsection{Convex optimization for the unbiased conditional probabilities  with i.i.d. and uniform input marginal  probabilities}

   Recall that we would like to check if the boundaries of the information causality and the generalized Tsirelson bound agree or not.
    To achieve this, we may maximize the information gain $I$ with the joint probabilities $\Pr(A_{\vec{x}},B_{\vec{y}}|\vec{x},\vec{y})$ realized by quantum mechanics. Or, we may
find  the generalized Tsirelson bound and then evaluate the corresponding  information gain $I$ which can be compared with the bound of information causality.
      These two tasks are not equivalent but complementary.
       However, unlike the first task, the second task will be concave problem as known in \cite{Wehner,Acinl}.
        The only question in this case is if the corresponding information gain $I$ is monotonically related to the Bell-type functions or not.
         If yes, then finding the generalized Tsirelson bound is equivalent to maximizing the information gain $I$ in our communication schemes.     The answer is partially yes as we will show  this monotonic relation holds only for the unbiased conditional probabilities $\Pr(\beta|a_{i},b=i)$ with i.i.d. and uniform input marginal probabilities $\Pr(a_{i})$.

 The unbiased conditional probabilities $\Pr(\beta|a_{i},b=i)$ are symmetric and isotropic. This is defined as follows. One can construct a matrix $CP$ with the matrix elements
$CP_{j+1,k+1}=\Pr(\beta=k|a_{i}=j,b=i)$ with $j,k \in \{0,1,...,d-1\}$. If
all the rows of matrix $CP$ are permutation for each other and all
columns are also permutation for each other, the conditional
probabilities $\Pr(\beta|a_{i},b=i)$ are symmetric. Moreover, if the
symmetric conditional probabilities $\Pr(\beta|a_{i},b=i)$ for
different $i$ are the same, the conditional probabilities
$\Pr(\beta|a_{i},b=i)$ are isotropic.

      Assuming Alice's input is i.i.d., we have Shannon entropy $H(\beta|b=i)=\log_{2} d$.  As $\Pr(\beta|a_{i},b=i)$ are unbiased, they are symmetric so that $\Pr(\beta=t|a_{i}=j,b=i)=\frac{(d-1)\xi_{i}+1}{d}$ for $t=j$, and
$\Pr(\beta=t|a_{i}=j,b=i)=\frac{1-\xi_{i}}{d}$ for $t\neq j$. Thus,
the information gain $I$ becomes
\begin{eqnarray}
I=k
\log_{2}d+\sum_{i=0}^{k-1}[\frac{(d-1)\xi_{i}+1}{d}\log_{2}(\frac{(d-1)\xi_{i}+1}{d})+(1-\frac{(d-1)\xi_{i}}{d})\log_{2}(\frac{1-\xi_{i}}{d})].
\end{eqnarray}
Moreover, $\Pr(\beta|a_{i},b=i)$ are also
isotropic, therefore $\xi_{i}=\xi$ $\forall i$. For such a case the
 information gain $I$ can be further simplified to
\begin{eqnarray}\label{mutual Iso}
I=k [
\log_{2}d+\frac{(d-1)\xi+1}{d}\log_{2}(\frac{(d-1)\xi+1}{d})+(1-\frac{(d-1)\xi}{d})\log_{2}(\frac{1-\xi}{d})].
\end{eqnarray}
The value of $\xi$ is in the interval $[0,1]$.
As $\xi$ is the noise parameter of the
channel with input $a_{i}$ and output $\beta$, then $\xi=0$
for the completely random channel  and $\xi=1$ for the noiseless
one, i.e., $\Pr(\beta=t|a_{i}=j,b=i)=\frac{1}{d}$ for $\xi=0$ and
$\Pr(\beta=t|a_{i}=t,b=i)=1$ for $\xi=1$.

  We can show that the information gain $I$ is monotonically increasing with the Bell-type functions parameterized by the noise parameter $\xi$.
   To see this, we calculate the first and second derivative of $I$ with respect to $\xi$ and obtain
\begin{eqnarray}
&&\frac{dI}{d\xi}=\frac{d-1}{d}\log\frac{(d-1)\xi+1}{1-\xi}, \notag\\
&&\frac{d^{2}I}{d\xi^{2}}=\frac{d-1}{d}(\frac{d-1}{(d-1)\xi+1}+\frac{1}{1-\xi}).\notag
\end{eqnarray}
From the above, we see that $\frac{dI}{d\xi}$ is always positive for
$\xi\in [0,1]$. Moreover, it is easy to see that $I$ is minimal at
$\xi=0$ since $\frac{d^{2}I}{d\xi^{2}}=d-1>0$. Thus, if the RAC
protocol has i.i.d. and uniform input marginal probabilities, the  information gain $I$ is a monotonically increasing function of $\xi$ for the the unbiased conditional
probabilities $\Pr(\beta|a_{i},b=i)$.


  \section{Finding the quantum violation of the Bell-type inequalities from the hierarchical semi-definite programming}\label{sec4}

 We now will prepare for numerically evaluating the maximum of the Bell-type function
\be\label{bellf}
 \sum_{\vec{y}} \xi_{\vec{y}} \qquad \mbox{with  $\xi_{\vec{y}}$ given in \eq{xi} \; and \; $\Pr(a_i)={1\over d}, \; \forall a_i, i$.}
\ee
It is monotonic increasing with   information gain $I$ under some assumptions.
  In order to ensure that the maximum of (\ref{bellf}) can be obtained by quantum resource, we have to use the same method as in \cite{Acin,Acinl}.  In \cite{Acin,Acinl}, they checked if a given set of probabilities can be reproduced from quantum mechanics or not. This task can be formulated as solving a hierarchy of semidefinite programming (SDP).

\subsection{Projection operators with quantum behaviors}

We will now briefly review the basic ideas in  \cite{Acin,Acinl} and
then explain how to use it for our program.  In \cite{Acin,Acinl} they use the projection operators for the following measurement scenario. Two distant partite Alice and Bob share a NS-box. Alice and Bob input $X$ and $Y$ to the NS-box, respectively, and obtain the corresponding outputs $a\in A$ and $b\in B$. Here $A$ and $B$ are used to denote the set of all possible Alice's and Bob's measurement outcomes, respectively. We use $X(a)$ and $Y(b)$ to denote the corresponding inputs. These outcomes can be associated with some sets of projection operators $\{E_{a}:a\in A\}$ and  $\{E_{b}:b\in B\}$.
      The joint probabilities of the NS-box can then be determined by the quantum state $\rho$ of the NS-box and the projection operators as following:
\begin{eqnarray}\label{qu}
\Pr(a,b)=\Tr(E_{a}E_{b}\rho).
\end{eqnarray}
Note that $\Pr(a,b)$ is the abbreviation of
$\Pr(A_{\vec{x}},B_{\vec{y}}|\vec{x},\vec{y})=\Tr(E_{A_{\vec{x}}}
E_{B_{\vec{y}}}\rho)$ defined in the previous sections.

   If  $E_{a}$ and $E_{b}$ are the genuine quantum operators,
    then they shall satisfy (i)  hermiticity: $E^{\dag}_{a}=E_{a}$ and $E^{\dag}_{b}=E_{b}$; (ii) orthogonality: $E_{a}E_{a'}=\delta_{aa'}$ if $X(a)=X(a')$ and $E_{b}E_{b'}=\delta_{b,b'}$ if $Y(b)=Y(b')$; (iii) completeness: $\Sigma_{ a\in X }E_{a}=\mathbb{I}$ and $\Sigma_{ b\in Y}E_{b}=\mathbb{I}$; and  (iv) commutativity: $[E_{a},E_{b}]=0$.

   In our measurement scenario,
    the distant partite Alice and Bob perform local measurements so that property (iv) holds.
     On the other hand, the property (iii) implies no-signaling as it leads to \eq{ns-box} via \eq{qu}.
      Furthermore, this property also implies that there is redundancy in specifying Alice's operators $E_a$'s with the same input since one of them can be expressed by the others.
 Thus, we can eliminate one of the outcomes per setting and denote the corresponding sets of the remaining outcomes for the input $X$ by $\tilde{A}_{X}$ (or $\tilde{B}_{Y}$ for Bob's outcomes with input $Y$).
        The collection of such measurement outcomes $\bigoplus_{X}\tilde{A}_{X}$ is denoted as $\tilde{A}$. Similarly, we denote the collection of Bob's independent outcomes as $\tilde{B}$.

    Using the reduced set of projection operators  $\{ E_a: a \in \tilde{A} \}$ and $\{ E_b: b \in \tilde{B} \}$, we can construct a set of operators $\emph{O}=\{O_{1},O_{2},...,O_{i},...\}$.  Here $O_{i}$ is some linear function of products of operators in $\{\mathbb{I} \cup \{E_{a}:a\in\tilde{A}\} \cup \{E_{b}:b\in\tilde{B}\}\}$.
The set $\emph{O}$ is characterized by a matrix $\Gamma$ given by
\begin{equation}\label{ga}
\Gamma_{ij}=\Tr(O^{\dag}_{i}O_{j}\rho).
\end{equation}
By construction, $\Gamma$ is non-negative definite, i.e.,
\begin{equation}\label{ncon2}
\Gamma\succeq 0.
\end{equation}
This can be easily proved as follows. For any vector
$v\in\mathbb{C}^{n}$ (assuming $\Gamma$ is a $n$ by $n$ matrix), one
can have
\begin{equation}
v^{\dag}\Gamma
v=\Sigma_{s,t}v^{*}_{s}\Tr(O^{\dag}_{s}O_{t}\rho)v_{t}=\Tr(V^{\dag}V\rho)\geq
0.
\end{equation}

Recall that our goal is to judge if a given set of joint
probabilities such as \eq{qu} can be reproduced by quantum mechanics
or not.
  In this prescription,  the joint probabilities are then encoded in the matrix $\Gamma$ satisfying the quantum constraints \eq{qu} and \eq{ncon2}.
However, $\Gamma$ contains more information than just joint
probabilities \eq{qu}.
 For examples, the terms appearing in the elements of $\Gamma$ such as $\Tr(E_{a}E_{a'}\rho), \Tr(E_{b}E_{b'}\rho)$ for $X(a)\neq X(a')$ and $Y(b)\neq Y(b')$ can not be expressed in terms of the joint probabilities of the NS-box.
  This is because these measurements are performed on the same partite (either Alice or Bob) and are not commutative. Therefore, to relate the joint probabilities of the NS-box to the matrix $\Gamma$, we need to find the proper combinations of $\Gamma_{ij}$ so that the final object can be expressed in terms of only the joint probabilities.  Therefore, given the joint probabilities, there shall exist some matrix functions $F_q$'s such that the matrix $\Gamma$ is constrained as follows:
\begin{equation}\label{ncon1}
\Sigma_{s,t}(F_{q})_{s,t}\Gamma_{s,t}=g_{q}
\end{equation}
where  $g_{q}$'s are the linear functions of joint probabilities
$\Pr(a,b)$'s.

  We then call the matrix $\Gamma$ a certificate if it satisfies \eq{ncon2} and \eq{ncon1} for a given set of joint probabilities of NS-box. The existence of the certificate will then be examined numerically by SDP. If the certificate does not exist, the joint probabilities cannot be reproduced by quantum mechanics.

    Examples on how to construct $F_q$ and $g_q$ for some specific NS-box protocols can be found in \cite{Acin,Acinl}. For illustration, here we will explicitly demonstrate the case not considered in \cite{Acin,Acinl}, that is the $k=2$, $d=3$ RAC protocol. We will use the notation which we defined in the previous sections. We start by defining the set of operators $\mathcal{E}=\{\mathcal{E}_{i}\}:=\mathbb{I}\cup\{E_{A_{x}}:A_{x}\in\{0,1\},x\in\{0,1,2\}\}\cup \{E_{B_{y}}:B_{y}\in\{0,1\},y\in\{0,1\}\}$ with the operator label $i\in\{0,1,2,...,m_{a},....,m_{a}+m_{b}\}$. The operator $\mathcal{E}_{i=0}$ is the identity operator $\mathbb{I}$, and $\mathcal{E}_{1<i\le m_{a}}\in E_{A_{x}}$, $\mathcal{E}_{m_{a}<i\le m_{a}+m_{b}}\in E_{B_{y}}$.

 The associated quantum constraints can be understood as the relations between joint probabilities $\Pr(a,b)$ and $\Tr(\mathcal{E}^{\dagger}_a \mathcal{E}_b \rho)$ (or marginal probabilities $\Pr(a)$ and $\Tr(\mathbb{I}\mathcal{E}_a \rho)$).  That is,
\begin{eqnarray}\label{d3k2con}
&&\Tr(\rho)=1, \qquad \Tr(\mathbb{I}E_{A_{x}}\rho)=\Pr(A_{x}|x),\qquad \Tr(\mathbb{I}E_{B_{y}}\rho)=\Pr(B_{y}|y),\notag\\
&&\Tr(E_{A_{x}}E_{A_{x}'}\rho)=\delta_{A_{x},A_{x}'}\Pr(A_{x}|x),\Tr(E_{B_{y}}E_{B_{y}'}\rho)=\delta_{B_{y},B_{y}'}\Pr(B_{y}|y),\notag\\
&&\Tr(E_{A_{x}}E_{B_{y}}\rho)=\Pr(A_{x},B_{y}|x,y).
\end{eqnarray}
Note that these equations also hold when permuting the operators, i.e.,
$\Tr(E_{A_{x}}E_{B_{y}}\rho)=\Tr(E_{B_{y}}E_{A_{x}}\rho)$.

   Moreover, we can make the matrix $\Gamma$ to be real and symmetric by redefining it as
$\Gamma=(\Gamma^{*}+\Gamma)/2$. Thus, in the following we will only
display the upper triangular part of $\Gamma$. We then use the
quantum constraints \eq{d3k2con} to construct $F_q$ and $g_q$ by
comparing them with \eq{ncon1}. We then see that every constraint in
\eq{d3k2con} yields a matrix function $F_q$ which has only one
non-zero element, and also yields a function $g_q$ which is either
zero or contains only a single term of a marginal or joint
probabilities. These constraints can be further divided into four
subsets labeled by $q=(q_{1},q_{2},q_{3},q_{4})$ as follows:

\begin{enumerate}

\item The labels $q_{1},q_{2}\in\{0,1,...,m_{a}+m_{b}\}$ are used to specify the marginal probabilities $\Tr(\mathbb{I}\mathcal{E}_{q_{1}}\rho)$ and $\Tr(\mathcal{E}^{\dagger}_{q_{2}}\mathcal{E}_{q_{2}}\rho)$. The corresponding matrix functions $F_q$ are given by $(F_{q_{1}})_{s,t}=\delta_{s,1}\delta_{t, q_{1}+1}$ and $(F_{q_{2}})_{s,t}=\delta_{s, q_{2}+1}\delta_{t, q_{2}+1}$, and the $g_{q_{1}}$ and $g_{q_{2}}$ are the corresponding marginal probabilities.

\item The label $q_{3}\in\{1,...,d^{k-1}+k\}$ is used to specify the probabilities associated with the orthogonal operator pairs, $\Tr(\mathcal{E}_{2q_{3}-1}\mathcal{E}_{2q_{3}}\rho)$. The matrix element $(F_{q_{3}})_{s,t}=\delta_{s, 2q_{3}}\delta_{t,2q_{3}+1}$, and $g_{q_{3}}=0$.

\item The label $q_{4}\in\{1,...,m_{a}m_{b}\}=4(2x+A_{x})+(2y+B_{y}+1)$ is used to specify the joint probabilities of the NS-box. The corresponding $F_q$ and $g_q$ are given by
$(F_{q_{4}})_{s,t}=\delta_{s, 2x+A_{x}+2}\delta_{t,
m_{a}+2y+B_{y}+2}$, and $g_{q_{4}}=\Pr(A_{x},B_{y}|x,y)$.

\end{enumerate}

Considering the above set of quantum constraint, we can define the
associated $\Gamma$ matrix
\begin{align}\label{d3k2n1}
 &\Gamma=\scriptsize\left(
             \begin{array}{ccccccccccc}
               1 & \Pr(0|0)_{A}&\Pr(1|0)_{A} & \Pr(0|1)_{A}&\Pr(1|1)_{A}&\Pr(0|2)_{A}&\Pr(1|2)_{A} & \Pr(0|0)_{B} & \Pr(1|0)_{B}& \Pr(0|1)_{B} & \Pr(1|1)_{B}\\
                 & \Pr(0|0)_{A}& 0         &  \chi_{0} &\chi_{1}  &\chi_{2}  &\chi_{3}   & \Pr(00|00)   & \Pr(01|00)  & \Pr(00|01)   &\Pr(01|01)   \\
                 &           &\Pr(1|0)_{A} & \chi_{4}  &\chi_{5}  &\chi_{6}  &\chi_{7}   & \Pr(10|00)   &\Pr(11|00)   &\Pr(10|01)    &\Pr(11|01)  \\
                 &           &             & \Pr(0|1)_{A}&  0     &\chi_{8}  &\chi_{9}   & \Pr(00|10)   &\Pr(01|10)   &\Pr(00|11)    &\Pr(01|11)  \\
                 &           &             &        & \Pr(1|1)_{A}&\chi_{10} &\chi_{11}  & \Pr(10|10)   &\Pr(11|10)   &\Pr(10|11)    &\Pr(11|11)  \\
                 &           &             &        &           &\Pr(0|2)_{A}&0          & \Pr(00|20)   &\Pr(01|20)   &\Pr(00|21)    &\Pr(01|21)  \\
                 &           &             &        &           &          & \Pr(1|2)_{A}& \Pr(10|20)   &\Pr(11|20)   &\Pr(10|21)    &\Pr(11|21)  \\
                 &           &             &        &           &          &           & \Pr(0|0)_{B} &0          &\chi_{12}    &\chi_{13} \\
                 &           &             &        &           &          &           &            &\Pr(1|0)_{B} &\chi_{14}    &\chi_{15} \\
                 &           &             &        &           &          &           &            &           &\Pr(0|1)_{B}    &0 \\
                 &           &             &        &           &          &           &            &           &0&\Pr(1|1)_{B} \\
             \end{array}
           \right),
\end{align}
where $\Pr(A_{x}|x)_{A}$'s and $\Pr(B_{y}|y)_{B}$'s are the
marginal probabilities for Alice and Bob, respectively,  and
$\Pr(A_{x},B_{y}|x,y)$'s are the joint probabilities of the NS-box.
The elements $\chi_i$'s in the above cannot be defined by the given
marginal and joint probabilities because they correspond to the
probabilities of different measurement settings for only one party.
Thus, they cannot appear in the constraints \eq{ncon1} but are still
constrained by the non-negative definiteness of $\Gamma$.

{\it Testing the existence of the certificate---} The task of
testing the existence of the certificate can be formulated as a SDP
by defining the standard primal and the associated dual problems.
 The details can be found in Appendix \ref{sdp}.
  The primal problem of SDP is subjected to certain conditions associated with a positive semi-definite matrix, which can be either linear equalities or inequalities.
    Each primal problem has an equivalent dual problem.
     Therefore, when the optimal value of the primal problem is the same as the optimal value of the dual problem, the feasible solution of the problem is obtained.

 For our case the primal problem of SDP is as follows:
 \begin{subequations}
\begin{align}
maximize \qquad & \lambda\\
subject \quad to \qquad    & \Tr(F_{q}^{T}\Gamma)=g_{q}, \qquad q=1,...,m, \\
         & \Gamma-\lambda \mathbb{I} \succeq 0.
\end{align}
\end{subequations}
Obviously, if the maximal value $\lambda\ge 0$ is obtained, the
non-negative definiteness of $\Gamma$ is guaranteed under the
quantum constraints \eq{ncon2}.

On the other hand, the associated dual problem is given by
 \begin{subequations}
\begin{align}
maximize \qquad & \sum_{q}y_{q}g_{q},\\
subject \quad to \qquad    & \sum_{q}y_{q}F_{q}^{T}\succeq 0, \\
         & \sum_{q}y_{q}\Tr(F_{q}^{T})=1.
\end{align}
\end{subequations}
Note that the quantity $\sum_{q}y_{q}g_{q}$ is the Bell-type
function since $g_q$'s are mainly the two-point correlation
function. Therefore, maximizing this quantity is equivalent to
finding the generalized Tsireslon bound. That is, if the solution
of this SDP is feasible, then the associated certificate exists and
there yields the generalized Tsireslon bound.

\subsection{Hierarchy of the semi-definite programming}

 Different operator sets $\emph{O}$'s yield different quantum constrains \eq{qu} and \eq{ncon2}.  There seems no guideline in choosing the set $\emph{O}$ and examining the existence of the corresponding certificate.  However, it is easy to see that the certificates associated with different operator sets are equivalent. This can be seen as follows.    Let us assume $\emph{O}$ and $\emph{O}'$ are two linearly equivalent set of operators such that  $O_{i}\in \emph{O}$ can be expressed by a linear combination of the elements in $\emph{O}'$, i.e., $O_{i}=\sum_{j}C_{i,j}O_{j}^{'}$. If there exists a matrix $\Gamma'$ satisfying \eq{ncon2} and\eq{ncon1} for the corresponding operator set $\emph{O}'$, then there will exist another matrix $\Gamma$ whose elements $\Gamma_{s,t}=\sum_{q,l}C_{q,s}^{*}\Gamma_{q,l}^{'}C_{l,t}$ are also satisfying \eq{ncon2} and \eq{ncon1} for the set $\emph{O}$.  Therefore, we only need to stick to one set of operators in this linear equivalence class when examining the existence of the corresponding certificate.

 Besides, a systematic way of constructing $\emph{O}$ is proposed in \cite{Acin,Acinl} so that the task of finding the certificate can be formulated as solving a hierarchy of SDP.  This is constructed as follows.  The length  of the operator $O_{i}$, denoted by $|\emph{O}_i|$, is defined as the minimal number of projectors used to construct it. We can then divide the set $\emph{O}$ into different subsets labeled by the maximal length of the operators in the corresponding subset. Thus, we decompose the operator set $\emph{O}$ into a sequence of hierarchical operator sets denoted by $S_n$ where $n$ is the maximal length of the operators in $S_n$.  That is,
\begin{eqnarray}
&&S_{0}=\{{\mathbb{I}}\}\notag\\
&&S_{1}=\{S_{0}\}\cup \{E_{a}:a\in\tilde{A}\} \cup \{E_{b}:b\in\tilde{B}\}\notag \\
&&S_{2}=\{S_{0}\}\cup \{S_{1}\} \cup
\{E_{a}E_{a'}:a,a'\in\tilde{A}\} \cup
\{E_{b}E_{b'}:b,b'\in\tilde{B}\}
\cup \notag\\
&&\{E_{a}E_{b}:a\in\tilde{A},b\in\tilde{B}\}\notag\\
...
\end{eqnarray}

 Furthermore, to save the computer memory space used in the numerical SDP algorithm,
  in the above sequence we can add an intermediate set  between $S_{n}$ and $S_{n+1}$,
   which is given by $S_{n+AB}:=\{S_{n}\}\cup \{S\in S_{n+1}| S=E_{a}E_{b}S':  a\in\tilde{A},b\in\tilde{B}\}$.
    For example, when $n=1$ we have $S_{1+AB}=\{S_{1}\}\cup\{E_{a}E_{b}:a\in\tilde{A},b\in\tilde{B}\}$ such that $S_{1}\subseteq S_{1+AB}\subseteq S_{2}$.
     Note that $S_{1+AB}$ doesn't have the product of the marginal projection operators in the form of $\{E_{a}E_{a'}:a,a'\in\tilde{A}\}$ and $\{E_{b}E_{b'}:b,b'\in\tilde{B}\}$.
      It is clear that $S_{1+AB}\subseteq S_{2}$.
        All the operators in $\emph{O}$ can be expressed in terms of the linear combination of the operators in $S_n$ for large enough $n$.

\begin{figure}[th]
\includegraphics[width=0.8\columnwidth]{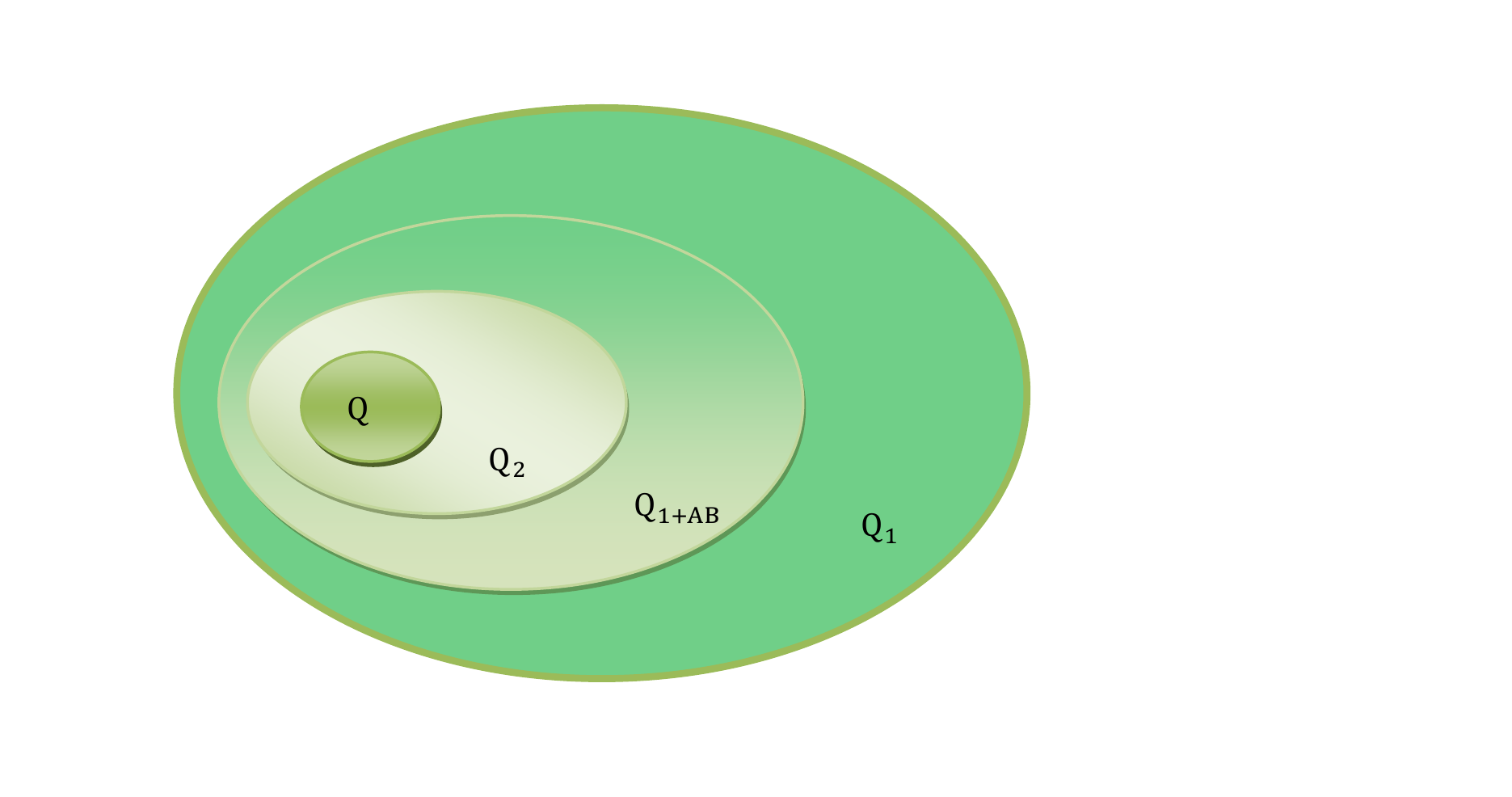}\newline
\caption{The geometric interpretation of collection $Q_{n}$}
\label{Qn}
\end{figure}

 Since we know $S_{n}\subseteq S_{n+AB}\subseteq S_{n+1}$, the associated constraints produced by $S_{n+1}$ is stronger than $S_{n+AB}$ and $S_{n}$.
  We can start the task from
$S_{1}$ then $S_{1+AB}$, $S_{2}$ and so on. Let the certificate
matrix associated with the set $S_n$ be denoted as $\Gamma^{(n)}$.
 Finding the certificate associated with this sequence can be formulated as a hierarchical SDP.
   Once the given joint probabilities satisfy the quantum constraints \eq{ncon2} so that the associated certificate $\Gamma^{(n)}$ exists,
    we then denote the collection of these joint probabilities as $Q_{n}$.
     Since we know that the associated constraints are stronger than the previous steps of the hierarchical sequence, the collection $Q_{n}$ will become smaller for the higher $n$. That is, the non-quantum correlations will definitely fail the test at some step in the hierarchical SDP.  The geometrical interpretation of the above fact is depicted in  Fig \ref{Qn}.

  It was shown in \cite{Acin, Acinl} that the probability is ensured to be quantum only when the certificate associated with $S_{n\rightarrow \infty}$ exists,
   i.e., for the joint probabilities in the collection $Q$ of Fig \ref{Qn}. In this sense, it seems that we have to check infinite steps. To cure this,
     a stopping criterion is proposed in \cite{Acin,Acinl} to terminate the check process at some step of the hierarchical SDP.
      This can ensure that the given joint probabilities are quantum at finite $n$ if the stopping criterion is satisfied.

The stopping criterion is satisfied when the rank of sub-matrix of
$\Gamma^{(n)}$ is equal to the rank of $\Gamma^{(n)}$, i.e.,
\begin{eqnarray} rank(\Gamma_{X,Y}^{(n)})=rank(\Gamma^{(n)}).
\end{eqnarray}
The element of $\Gamma_{X,Y}^{(n)}$ is constructed by the operators
 in the set $S_{X,Y}:=\{S_{n-1}\}\bigcup \{S=E_{a}E_{b}S': a\in \tilde{A}_{X}, b\in \tilde{B}_{Y}, |S|\leq n\}$.

The above stopping criterion is for integer $n$. However, it was
also generalized in \cite{Acinl} for the intermediate certificate
$\Gamma^{(n+AB)}$: the stopping criterion is satisfied if the
following equation is satisfied for all the measurement settings $X$
and $Y$,
\begin{eqnarray}\label{rank}
rank(\Gamma^{(n+XY)})=rank(\Gamma^{(n+AB)}),
\end{eqnarray}
so that the certificate $\Gamma^{(n+AB)}$ has a rank loop. Here
$\Gamma^{(n+XY)}$ is the certificate associated with
  $S_{n+XY}:=\{S_{n}\}\cup \{S\in S_{n+1}| S=E_{a}E_{b}S': a\in \tilde{A}_{X}, b\in \tilde{B}_{Y}\}$.

   Now we are ready to implement the above criterion to numerically examine the quantum behaviors of the given joint probabilities for our RAC protocols with higher $k$ and $d$.

\subsection{The quantum violation of the Bell-type inequalities and the corresponding information gain in the hierarchical semi-definite programming}

Any Bell-type function including (\ref{bellf}) can be written as the
linear combination of joint probabilities, then the hierarchical SDP
can be used to approach the quantum bound of the Bell-type functions
(the generalized Tsirelson bound). Recall that the value of the
Bell-type functions and the information gain $I$ are monotonically
related for  the unbiased conditional
probabilities $\Pr(\beta|a_{i},b=i)$ with i.i.d. and uniform input
marginal probabilities $\Pr(a_{i})$. After obtaining the maximum of
the Bell-type functions at each step of the aforementioned
hierarchical SDP,
 we can calculate the corresponding information gain $I$ and compare with the information causality.
 Since the quantum constraint is stronger in the hierarchical SDP and the collection of $Q_{n}$ will become smaller while $n$ is increasing. We then know that the bound of the Bell-type functions and the associated  information gain $I$ will become tighter for larger $n$ and it will converge to the quantum bound for large enough $n$. Once the bound of   information gain $I$ at some step of hierarchy doesn't saturate the information causality, we can then infer that  the quantum bound of   information gain will not saturate the information causality, too.

First, let us discuss how to find
the generalized Tsirelson bound of
the Bell-type functions. As discussed before,
 the problem of finding the generalized Tsirelson bound can be reformulated as a SDP. The primal problem of
this SDP is defined as
\begin{subequations}\label{P1}
\begin{align}
maximize \qquad &  \Tr(C^{T}\Gamma^{(n)})\\
subject \quad to \qquad &  \Tr(F_{q}^{T}\Gamma^{(n)})=g_{q}(p), \quad q=1, \cdots, m;\label{sdpc1}\\
                        &\Gamma^{(n)}\succeq 0. \label{sdpc2}\\
                        &\Tr(H_{w}^{T}\Gamma^{(1)})\geq 0,\quad w=1,\cdots,s;\label{sdpc3}
                        \end{align}
\end{subequations}
    The matrix $C$ is given to make $\Tr(C^{T}\Gamma^{(n)})$ the Bell-type functions which we would like to maximize. Eq. \eq{sdpc1} and \eq{sdpc2} are the quantum constraints discussed in the previous subsections so that the quantum behaviors are ensured during the SDP procedure. Moreover, with proper choice of the matrix $H_w$ \footnote{ Since we only consider $a\in\tilde{A}$ and $b\in\tilde{B}$ to save the computer memory space, we need to choose $H_w$ to ensure the non-negative definiteness of not only the $(d-1)^2$ terms of $\Gamma^{(1)}$ but also the other $d^{2}-(d-1)^{2}$ terms which are the linear combinations of the elements of $\Gamma^{(1)}$.}, the condition \eq{sdpc3} is introduced to ensure the non-negativity of the joint probabilities which are the off-diagonal elements of $\Gamma^{(1)}$.

 In the following we define the matrix $C$ for our case. Eq. \eq{bellf}, which can be expressed as the linear combination of the joint probabilities, i.e., $\sum_{\vec{x},\vec{y}}\Pr(B_{\vec{y}}-A_{\vec{x}}=\vec{x}\cdot\vec{y}|\vec{x},\vec{y})$, is the object for our SDP \eq{P1}.  Since we only consider $d-1$ marginal probabilities per measurement setting, we should further rewrite our object according to the completeness conditions, i.e., $\Sigma_{ a\in X }E_{a}=\mathbb{I}$ and $\Sigma_{ b\in Y}E_{b}=\mathbb{I}$. After rewriting, we can write down the matrix
$C$ in \eq{P1}. We take $d=3$, $k=2$ RACs protocol for example. For
 $\Gamma^{(1)}$,
\begin{align}\label{cd3k2n1}
&C=\frac{1}{2}\left(\begin{array}{ccccccccccc}
               1.&  2. & 0.&  0.& -1.&  1.&  1.&  0.&  3.&  0.& 0.\\
              2. & 0.&  0.&  0.&  0.&  0.&  0.& -1.& -2.& -1.& -2.  \\
              0. & 0.&  0.&  0.&  0.&  0.&  0.&  1.& -1.&  1.& -1.  \\
               0. & 0.&  0.&  0.&  0.&  0.&  0.& -1.& -2.&  2.&  1.  \\
              -1.&  0.&  0.&  0.&  0.&  0.&  0.&  1.& -1.&  1.&2.\\
              1.&  0.  &0.  &0.  &0.  &0.  &0. &-1. &-2. &-1. & 1.\\
             1.  &0. & 0. & 0.  &0.  &0.  &0.  &1. &-1. &-2. &-1.\\
             0.  &-1.  &1.  &-1.  &1.  &-1.  &1.  &0.  &0.  &0.  &0.\\
            3.  &-2.  &-1.  &-2.  &-1.  &-2.  &-1.  &0.  &0.  &0.  &0.\\
            0.  &-1.  &1.  &2.  &1.  &-1.  &-2.  &0.  &0.  &0.  &0.\\
            0.  &-2.  &-1.  &1.  &2.  &1.  &-1.  &0.  &0.  &0.  &0.\\
             \end{array}
           \right).
\end{align}
The size of \eq{cd3k2n1} is equal to the size of $\Gamma^{(1)}$ (the
first step in our hierarchical SDP). If $n\neq 1$, the size of
matrix $C$ will be bigger, we could define \eq{cd3k2n1} as the
sub-matrix of matrix $C$ and the other elements of $C$ are zero such
that the object functions $\Tr(C^{T}\Gamma^{(n)})$ are all equal for
different steps of our hierarchical SDP.

For higher $d$ and $k$, we write down the quantum constraints
\eq{ncon2} for $\Gamma^{(1)}$ and $\Gamma^{(1+AB)}$ and estimate its
number in Appendix \ref{higherSDP}. However, due to the limitation
of the computer memory (we have $128GB$), we cannot finish all the
tests of our hierarchical SDP but stop at level of $1+AB$.  In our
calculation, we take the
 $\sum_{\vec{x},\vec{y}}\Pr(B_{\vec{y}}-A_{\vec{x}}=\vec{x}\cdot\vec{y}|\vec{x},\vec{y})$ as the object of SDP, which is monotonically related to the Bell-type functions $\sum_{\vec{y}} \xi_{\vec{y}}$ in a straightforward way via \eq{xi}.

 At the $n=1$ level the numerical results of our SDP object $\sum_{\vec{x},\vec{y}}\Pr(B_{\vec{y}}-A_{\vec{x}}=\vec{x}\cdot\vec{y}|\vec{x},\vec{y})$  for various $k$ and $d$ are listed below:
\begin{center}
\begin{tabular}{|c|c|c|c|c|}
\hline
k        &d=2        &d=3   &d=4  &d=5            \\
\hline
 2  &3.4142   &4.8284&6.2426     &7.6569   \\
\hline
3   &9.4641    &19.3923& 32.7846 & 49.6410\\
\hline
4   &24.0000    &72.0000  &160.0000       &\\
\hline
5   &57.8885     & 255.7477  &       &\\
\hline
\end{tabular}
\end{center}
The entries in the table are the values of
$\sum_{\vec{x},\vec{y}}\Pr(B_{\vec{y}}-A_{\vec{x}}=\vec{x}\cdot\vec{y}|\vec{x},\vec{y})$.

Similarly, at the $n=1+AB$ level the results for the same SDP object
are listed below:
\begin{center}
\begin{tabular}{|c|c|c|c|c|}
\hline
k        &d=2        &d=3   &d=4  &d=5            \\
\hline
 2  &3.4142   &4.6667&5.9530 &7.1789    \\
\hline
3   &9.4641    &18.6633&      & \\
\hline
4   &24.0000   &     &       &\\
\hline
5   &57.8885    &     &       &\\
\hline
\end{tabular}
\end{center}

The stopping criterion is checked at the same time. Unfortunately,
it is not satisfied for $\Gamma^{(1+AB)}$,
 this means that the bound associated with $\Gamma^{1+AB}$ is not the generalized Tsirelson bound. However,
  our numerical computational capacity cannot afford for the higher level calculations.

  Few more remarks are in order: (i) Even we do not require   $\Pr(\beta|a_{i},b=i)$ to be isotropic, i.e., uniform $\xi_{\vec{y}}$ for our SDP, the final results show that the $\Pr(\beta|a_{i},b=i)$'s  maximizing the SDP object are isotropic for our level $n=1$ and $n=1+AB$ check.
      (ii) We find the bound at the $n=1$ level is the same as the bound derived from the signal decay theorem in section \ref{sec2}.
       (iii) For $d=2$ case, the bound for the SDP object at the $n=1$ and $n=1+AB$ level are equal,
        which is also the same as the Tsirelson bound as gurantteed by Tsirelson's theorem \cite{Wehner}.
         Since the bound is already the Tsirelson bound, it will not change for the further steps of the hierarchical SDP.
           (iv) For $d>2$, the bound of the SDP object at the $n=1+AB$ level becomes tighter than the one at the $n=1$ level,
            as expected. However, it needs more numerical efforts to arrive the true tight bound for the quantum violation of the Bell-type inequalities, i.e., the generalized Tsireslon
     bound.

Since  the conditional probabilities
$\Pr(\beta|a_{i},b=i)$ are unbiased for the above SDP procedure,
 we can then obtain the value of the noise parameter $\xi$ and use \eq{mutual Iso} to evaluate the corresponding   information gain $I$:

At the $n=1$ level,
\begin{center}
\begin{tabular}{|c|c|c|c|c|}
 \hline
        &d=2        &d=3   &d=4  &d=5            \\
\hline
 Information causality    & 1.0000  &1.5850&2.0000&2.3220\\
\hline
 k=2  &0.7982 & 1.3547&  1.7845 & 2.1357    \\
\hline
k=3   &0.7680 & 1.3360 & 1.7895 & 2.1680\\
\hline
k=4   &0.7549 &1.3333  &1.8048&  \\
\hline
k=5   &0.7476 &1.3345  &      &\\
\hline
\end{tabular}
\end{center}
The entries are the corresponding   information gain $I$ given by
\eq{mutual Iso}.

At the $n=1+AB$ level,
\begin{center}
\begin{tabular}{|c|c|c|c|c|}
\hline
        &d=2        &d=3   &d=4  &d=5            \\
\hline
 Information causality     & 1.0000   &1.5850&2.0000&2.3220\\
\hline
 k=2  &0.7982&  1.1972&  1.5478 & 1.7788\\
\hline
 k=3   &0.7680 & 1.1531&      & \\
\hline
k=4   &0.7549  &     &       &\\
\hline
k=5   &0.7476    &     &       &\\
\hline
\end{tabular}
\end{center}

\bigskip

Note that our results support the information causality.
 This is because the maximal   information gain $I$ evaluated from the joint probabilities constrained by the $n=1$ certificates is already smaller than the bound from the information causality.
  Thus, as implied by the geometric picture of Fig. \ref{Qn}, the the quantum bound on the  information gain $I$ obtained in the large $n$ limit will also satisfy the  information causality,
   at least for  the unbiased conditional probabilities with i.i.d. and uniform input marginal probabilities.
    Moreover, for a given $d$  the maximal  information gain $I$ from
the certificates decreases as $k$ increases. However, it is hard to
find the quantum bound of the  information gain $I$ exactly because
the stopping criterion fails at the $n=1+AB$ level. It needs more
checks for higher $n$ certificate to arrive the quantum bound of the
 information gain $I$. However, we will not carry out this task due
to the limitation of the computational power.

\section{Maximizing   information gain for general conditional probabilities  realized by quantum mechanics}\label{sec5}

Most of the RACs protocols discussed so far and in the literatures
are under some assumptions such as i.i.d., uniform input marginal
probabilities $\Pr(a_{i})$ for the
unbiased conditional probabilities $\Pr(\beta|a_{i},b=i)$. If we
want to test the information causality for more general cases, we
should try to find the maximum of the information gain $I$ for
the more general $\Pr(a_{i})$ and $\Pr(\beta|a_{i},b=i)$ but which can still be realized  quantum mechanically.

The conditional probabilities
$\Pr(\beta|a_{i},b=i)$ are the functions of the input marginal
probabilities $\Pr(a_{i})$ and the joint probabilities of NS-box.
Recall that from the proof of section \ref{sec3}, we cannot
formulate the problem of maximizing the  information gain $I$ as a convex optimization programming over Alice's input marginal probabilities and the joint probabilities of the NS-box.
  Thus,  for the case with the more general
conditional probabilities $\Pr(\beta|a_{i},b=i)$ but which can still be realized quantum
mechanically, we are forced to solve the problem by brutal force. The
procedure is as follows. Firstly, we divide the defining domains of
the joint and Alice's input marginal probabilities into many fine
points. We then pick up the points satisfying the consistent
relations for  the given conditional
probabilities $\Pr(\beta|a_{i},b=i)$. Secondly, we test if these
joint probabilities can be reproduced by quantum mechanics or not.
   If they can, we then evaluate the corresponding   information gain $I$.
     Thirdly, by comparing these  information gain $I$'s, we can obtain the maximal one and then check if the information causality is satisfied or not.
        By this brute-force method, we can then obtain the distribution of  information gain $I$ over the joint and the Alice's input marginal probabilities produced by quantum mechanics.
         This yields far more than just the maximal information gain consistent with quantum mechanics.
            The price to pay is the cost for the longer computing time.
              Due to the restriction of the computer power, we can only work for $d=2$ and $k=2$ case.

We start the discussion for  the case
with the more general conditional probabilities
$\Pr(\beta|a_{i},b=i)$ by fixing either the joint probabilities
$\Pr(B_{y}-A_{x}|x,y)$ or the input marginal probabilities
$\Pr(a_{i})$. Firstly, we assume the input probabilities are i.i.d.
and uniform such that we could take the CHSH function as the
Bell-type function. Therefore we could study the relation between
the   information gain $I$ and the quantum violation of the
Bell-type inequalities. Note that, when requiring
conditional probabilities
$\Pr(\beta|a_{i},b=i)$ \eq{d200channel} to have the i.i.d. and
uniform input marginal probabilities $\Pr(a_{i})$,
 the conditional probabilities
$\Pr(\beta|a_{i},b=i)$ then becomes symmetric automatically.
Secondly, in order to study the influence of the input marginal
probabilities $\Pr(a_{i})$ on the  information gain $I$, we pick up
three sets of the joint probabilities $\Pr(B_{y}-A_{x}|x,y)$
constrained by quantum mechanics and then evaluate the corresponding
 information gain with different input marginal probabilities
$\Pr(a_{i})$. Besides these conditional probabilities
$\Pr(\beta|a_{i},b=i)$, in order to test if the information
causality is always satisfied, we will consider
the case with the most general
conditional probabilities $\Pr(\beta|a_{i},b=i)$ but which can still be realized quantum
mechanically. Namely, we do not impose any condition on the
 conditional probabilities
$\Pr(\beta|a_{i},b=i)$ except the quantum constraints for the joint
probabilities of the NS-box.

Before evaluating the corresponding  information gain,
 the chosen joint probabilities $\Pr(B_{y}-A_{x}|x,y)$ should pass a test. For $d=2$ and $k=2$ RAC protocol, the quantum constraint is as follows:
\begin{eqnarray}\label{d2k2con}
 G=\left(
  \begin{array}{cccc}
    1 & \theta_{1}&C_{00}&C_{01} \\
      & 1&C_{10}&C_{11} \\
        & &1&\theta_{2} \\
          &  & &1 \\
  \end{array}
\right)\succeq 0,
\end{eqnarray}
where $C_{x,y}:=(-1)^{xy}[2 \Pr(B_{y}-A_{x}=x y|x,y)-1]$ is the
correlation function of the measurement setting $x$, $y$ for Alice
and Bob,
 respectively.
  The condition was pointed out in \cite{Wehner,Landau,Acin,Acinl}
   and can be derived as the necessary and sufficient condition for the quantum correlation functions $C_{x,y}$
    (or equivalently the joint probabilities $\Pr(B_{y}-A_{x}|x,y)$) by Tsirelson's theorem \cite{T3}, in which the marginal probabilities $\Pr(A_{x}|x)$ and $\Pr(B_{y}|y)$ are unbiased. Actually, $G$ is the sub-matrix of the $n=1$ certificate $\Gamma^{(1)}$. Due to the positivity, \eq{d2k2con} is satisfied once $\Gamma^{(1)}\succeq 0$.

Since the condition \eq{d2k2con} is related to a positive
semi-definite matrix, we need to use the numerical recipe to solve
it.
 Once the joint probabilities are not fixed in the  conditional probabilities
$\Pr(\beta|a_{i},b=i)$, we have to pick up many sets of joint
probabilities from their defining domains. This seems not efficient
enough to test all possible sets of joint probabilities by SDP.
Therefore, instead of using condition \eq{d2k2con} we use a set of linear
inequalities to test if the joint probabilities can be produced by
quantum mechanics or not. In this way, the test will become simpler
and more efficient. The linear inequalities are \cite{masa,T2}
\bes\label{masaness}
\begin{eqnarray}
|arcsin(C_{00})+arcsin(C_{01})+arcsin(C_{10})-arcsin(C_{11})|\leq
\pi,\\
|arcsin(C_{00})+arcsin(C_{01})-arcsin(C_{10})+arcsin(C_{11})|\leq
\pi,\\
|arcsin(C_{00})-arcsin(C_{01})+arcsin(C_{10})+arcsin(C_{11})|\leq
\pi,\\
|-arcsin(C_{00})+arcsin(C_{01})+arcsin(C_{10})+arcsin(C_{11})|\leq
\pi.
\end{eqnarray}
\ees Actually, the condition \eq{masaness} is equivalent to
\eq{d2k2con}. If the linear inequalities \eq{masaness} are
satisfied, then we can find valid $\theta_{1}$ and $\theta_{2}$ to
make condition \eq{d2k2con} satisfied, and vise versa
\cite{Landau,Acin,Acinl}.

 Once the corresponding correlation functions $C_{x,y}$ satisfy \eq{masaness}, we will know that these joint probabilities $\Pr(B_{y}-A_{x}|x, y)$ can be reproduced by quantum system.
But we have to notice that some of them could also be expressed by
the local hidden variable model. This means the shared correlation
is local. Since the bound of the CHSH function for local
correlations is different from the quantum non-local ones, we could
use the value of the CHSH function to divide them.

\subsection{Symmetric conditional probabilities with i.i.d. and uniform input marginal
probabilities}

 We start with the most simple case: the $d=2$, $k=2$ RAC protocol with the symmetric conditional probabilities
$\Pr(\beta|a_{i},b=i)$ and i.i.d., uniform input marginal
probabilities $\Pr(a_{i})$. In this case, the CHSH function
($|C_{0,0}+C_{0,1}+C_{1,0}-C_{1,1}|$) is equivalent to the Bell-type
function \eq{bellf}.
    Moreover, using the CHSH function and its three symmetric partners by shifting the minus sign, we could ensure that the shared correlations can be described by the local hidden variable model.
     Once the corresponding values of all these functions are less than $2$, the shared correlation is local.
     Otherwise, the shared correlation could be quantum non-local or
     beyond. The latter happens when some of these values are larger than $2\sqrt{2}$ which is nothing but the Tsirelson bound.
       When the Tsirelson bound is reached, the quantum violation of the CHSH inequality is the maximum.

\begin{figure}
\begin{minipage}[t]{1\linewidth}
\centering
\includegraphics[width=4in]{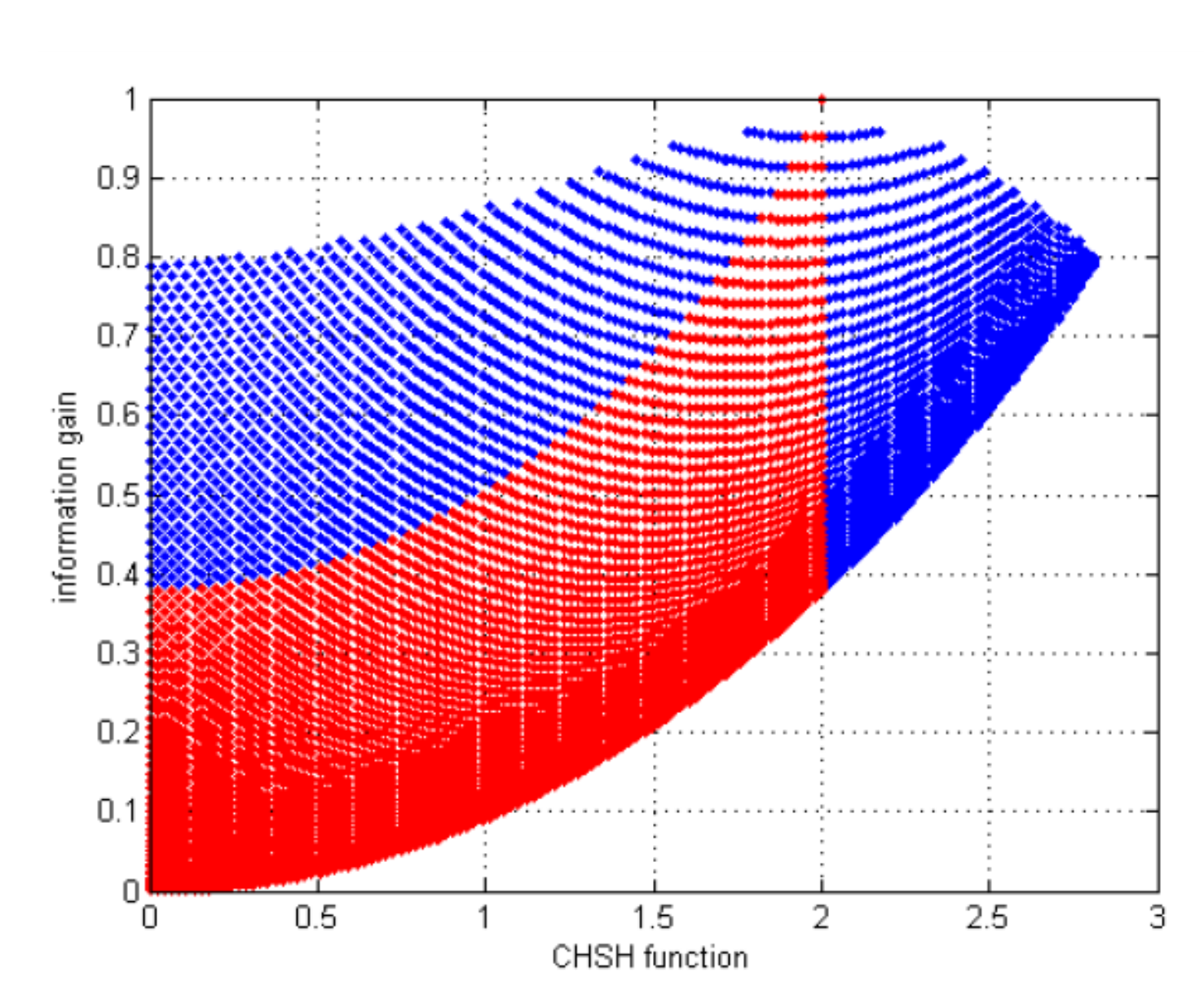}
\caption{Information gain v.s. the value of CHSH function for
$d=2$, $k=2$ RAC (quantum) protocol with i.i.d. and uniform input
marginal probabilities $\Pr(a_{i})$. The red part can be achieved
also by sharing the local correlation.} \label{d2k2g}
\end{minipage}%
\end{figure}
\begin{figure}
\begin{minipage}[t]{1\linewidth}
\centering
\includegraphics[width=4in]{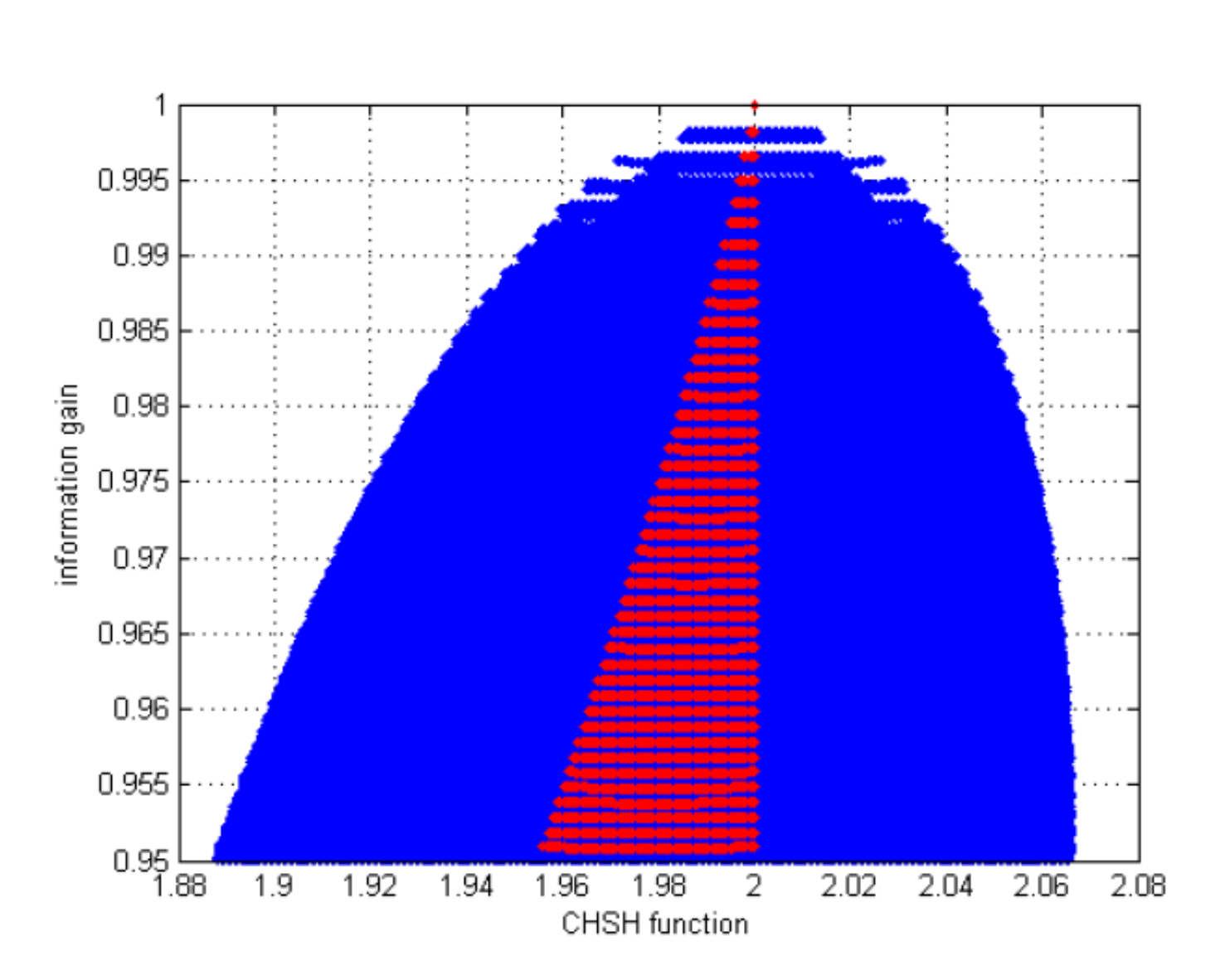}
\caption{Some points near the top region in Fig. \ref{d2k2g}.}
\label{d2k2n2}
\end{minipage}
\end{figure}

 In our numerical calculations,
  we divide the defining domain of the joint probabilities $\Pr(B_{y}-A_{x}|x, y)$ into $100$ points.
   Follow the procedure of our brute-force method, we obtain the distribution of the  information gain $I$ over the value of the CHSH function. The result is shown in Fig \ref{d2k2g}. for symmetric conditional probabilities
$\Pr(\beta|a_{i},b=i)$ with i.i.d. and uniform input marginal
probabilities $\Pr(a_{i})$. Note that, in Fig \ref{d2k2g}, all the
points satisfy quantum constraint \eq{masaness}. We particularly use
the red color to denote the points which also can be obtained by the
local correlations, i.e.,  the value
of the CHSH function and its three symmetric partners are all less
than $2$. Moreover, it seems that the distribution of the
information gain over the value of the CHSH function as shown in Fig
\ref{d2k2g} is not continuous. This is not the case but because we
did not partition the defining domain of the joint probabilities
fine enough.

          In Fig \ref{d2k2n2} we partition more finely on the defining domain of the joint probabilities in the top region of Fig \ref{d2k2g} and show that the empty region in Fig \ref{d2k2g} is now filled.
            Similarly, the empty region on the top of Fig \ref{d2k2n2} could be filled again by the more fine partitioning.

    The results in Fig \ref{d2k2g} is consistent with the information causality since the maximal  information gain for the local or quantum correlations is bound by 1, the bound suggested by information causality.
     However, the peculiar part of Fig \ref{d2k2g} is that some of the local correlations can achieve the larger information gain than $I\simeq 0.8$, which is achieved by the correlations saturating the Tsireslon bound.
      This peculiar part is the red region above $I\simeq 0.8$ in Fig \ref{d2k2g}.
      Especially, the maximal   information gain $I=1$ is reached when the shared correlation saturates the Bell inequality, i.e., the value of the CHSH function is equal to $2$. This indicates that the  information gain is not monotonically related to the CHSH function. Or
put this in the other way, the more amount of the quantum violation
of Bell-type inequalities may not always yield the more
information gain. We think it is interesting to understand this
phenomenon in the future works.

\begin{figure}[h]
\includegraphics[width=0.5\columnwidth]{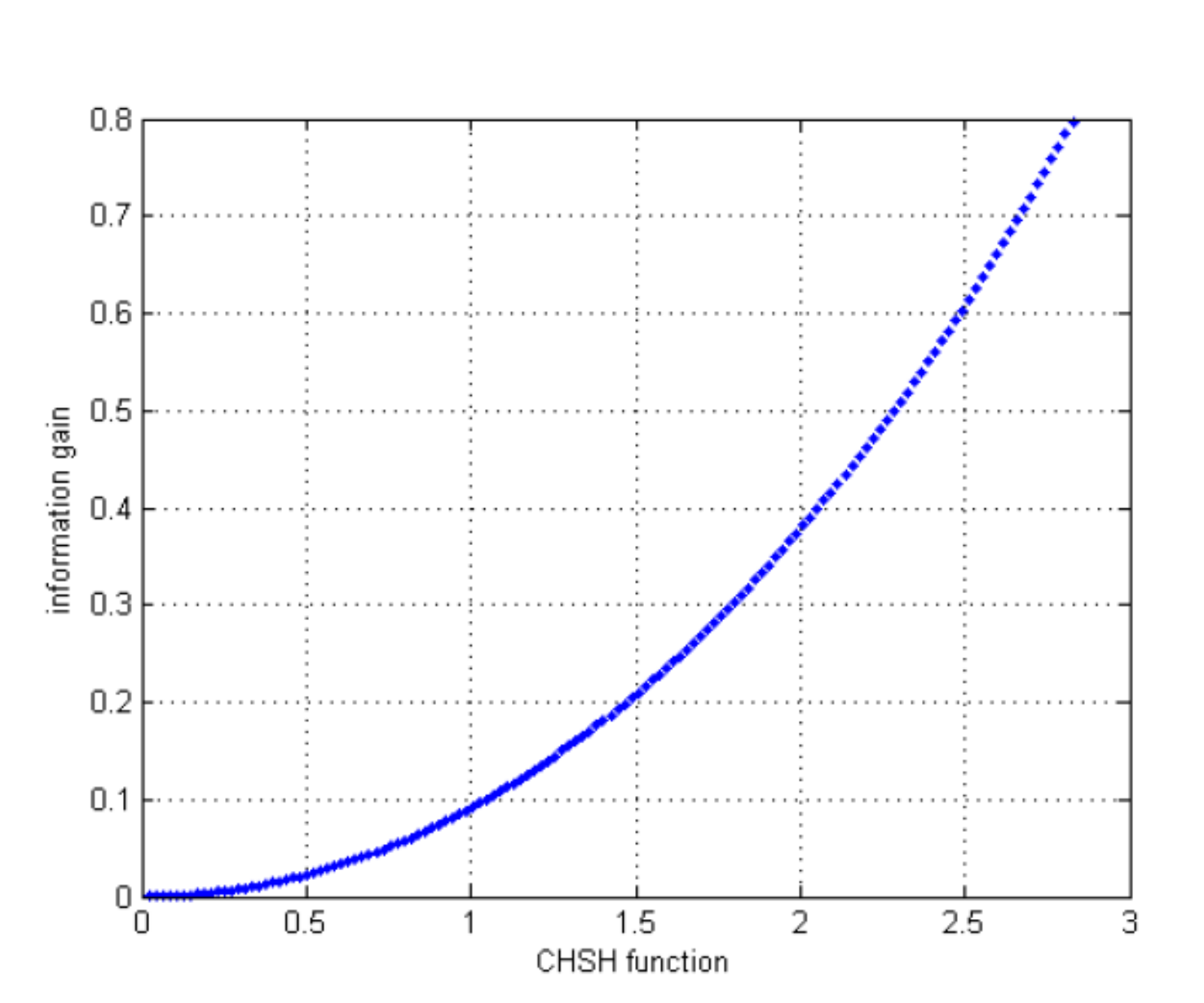}\newline
\caption{Information gain vs the value of CHSH function for
isotropic channels with i.i.d. and uniform input marginal
probabilities.} \label{d2k2iso}
\end{figure}

Form these symmetric conditional
probabilities $\Pr(\beta|a_{i},b=i)$ realized quantum mechanically
with i.i.d. and uniform input marginal probabilities $\Pr(a_{i})$,
we pick up the isotropic ones ($\xi_{0}=\xi_{1}$) and obtain Fig
\ref{d2k2iso}. It shows that the  information gain $I$ and the
value of the CHSH function achieved by quantum mechanics are
monotonically related. This explicitly demonstrate what we have
discussed in the previous section.

\subsection{Conditional probabilities with non-uniform input marginal probabilities}

In the above  conditional
probabilities $\Pr(\beta|a_{i},b=i)$, the input marginal
probabilities $\Pr(a_{i})$ are fixed to be i.i.d. and uniform
$\Pr(a_{i})$. Now we would like to demonstrate the effect of
non-uniform $\Pr(a_{i})$. In this case, we would like to vary
$\Pr(a_{i})$ but keep the joint probabilities of the NS-box fixed.
To see this effect for different conditional probabilities
$\Pr(\beta|a_{i},b=i)$, we consider three different sets of the
joint probabilities corresponding to (i) symmetric, (ii) symmetric
and isotropic and (iii) asymmetric
conditional probabilities $\Pr(\beta|a_{i},b=i)$.

   To be more specific,  for the case
   (i) the joint probabilities should be constrained by $\Pr(B_{y}-A_{x}=0|x,y=0)=1$ and $\Pr(B_{y}-A_{x}=xy|x,y=1)=\frac{1}{2}$ for $x=0,1$
    such that the noise parameters are given by $\xi_{0}=1$ and $\xi_{1}=0$. For the case (ii) all the joint probabilities $\Pr(B_{y}-A_{x}=xy|x,y)$ are equal to $\frac{1}{2}(1+\frac{1}{\sqrt{2}})$ such that $\xi_0=\xi_1=\frac{1}{\sqrt{2}}$. For the case (iii) the joint probabilities are given by $\Pr(B_{y}-A_{x}=0|x=0,y)=\frac{1}{2}(1+\frac{1}{\sqrt{2}})$ and
$\Pr(B_{y}-A_{x}=xy|x=1,y)=\frac{1}{2}$ for $y=0,1$. Obviously, it
is asymmetric for general input marginal probabilities $\Pr(a_{i})$.

  In the following discussion,
    we denote the mutual information $I(a_{0};\beta|b=0)$ as $I_{0}$ and $I(a_{1};\beta|b=1)$ as $I_{1}$,
     which are functions of two input marginal probabilities, namely, $\Pr(a_0=0)$ and $\Pr(a_1=0)$.
       Here $I_i$ can be thought as the mutual information between $a_i$ and $\beta$, and the corresponding noise parameter is $\xi_i$.
         The information gain $I$ is just $I=I_0+I_1$.
          Note that, $I_0$ does not depend on $\Pr(B_{y}-A_{x}=xy|x,y=1)$ and $I_1$ not on $\Pr(B_{y}-A_{x}=0|x,y=0)$.
           Thus, the  conditional probabilities
$\Pr(\beta|a_{0},b=0)$ for $I_0$ can be made symmetric by just
requiring $\Pr(B_{y}-A_{x}=xy|x,y=0)$'s for $x=0,1$ are equal, and
similarly for $\Pr(\beta|a_{1},b=1)$ for $I_1$ to be symmetric. An
important feature for these symmetric
conditional probabilities
$\Pr(\beta|a_{0},b=0)$ is that $I_i$ will depend only on $\Pr(a_i)$
not on $\Pr(a_{(i+1 \mod 2)})$.

\begin{figure}
\begin{minipage}[t]{0.5\linewidth}
\centering
\includegraphics[width=3in]{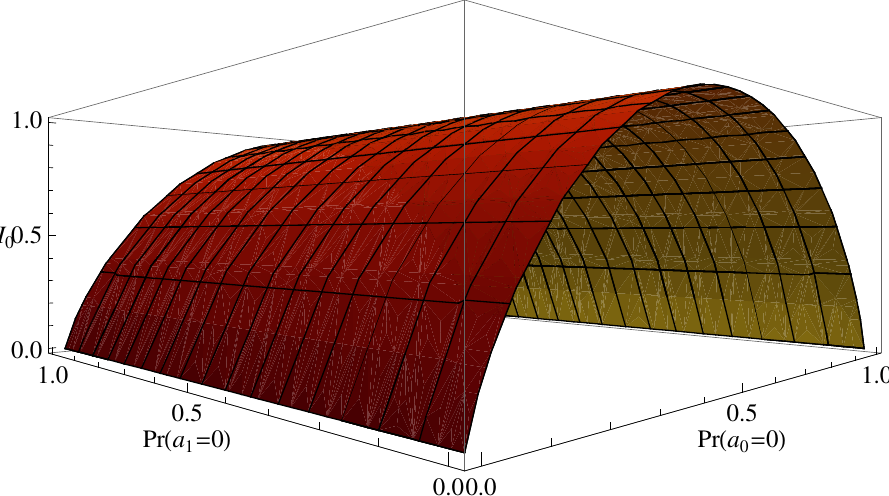}
\caption{$I=I_0$ vs $\Pr(a_{0,1}=0$) for case (i).} \label{Cij1I0}
\end{minipage}%
\begin{minipage}[t]{0.5\linewidth}
\centering
\includegraphics[width=2.3in]{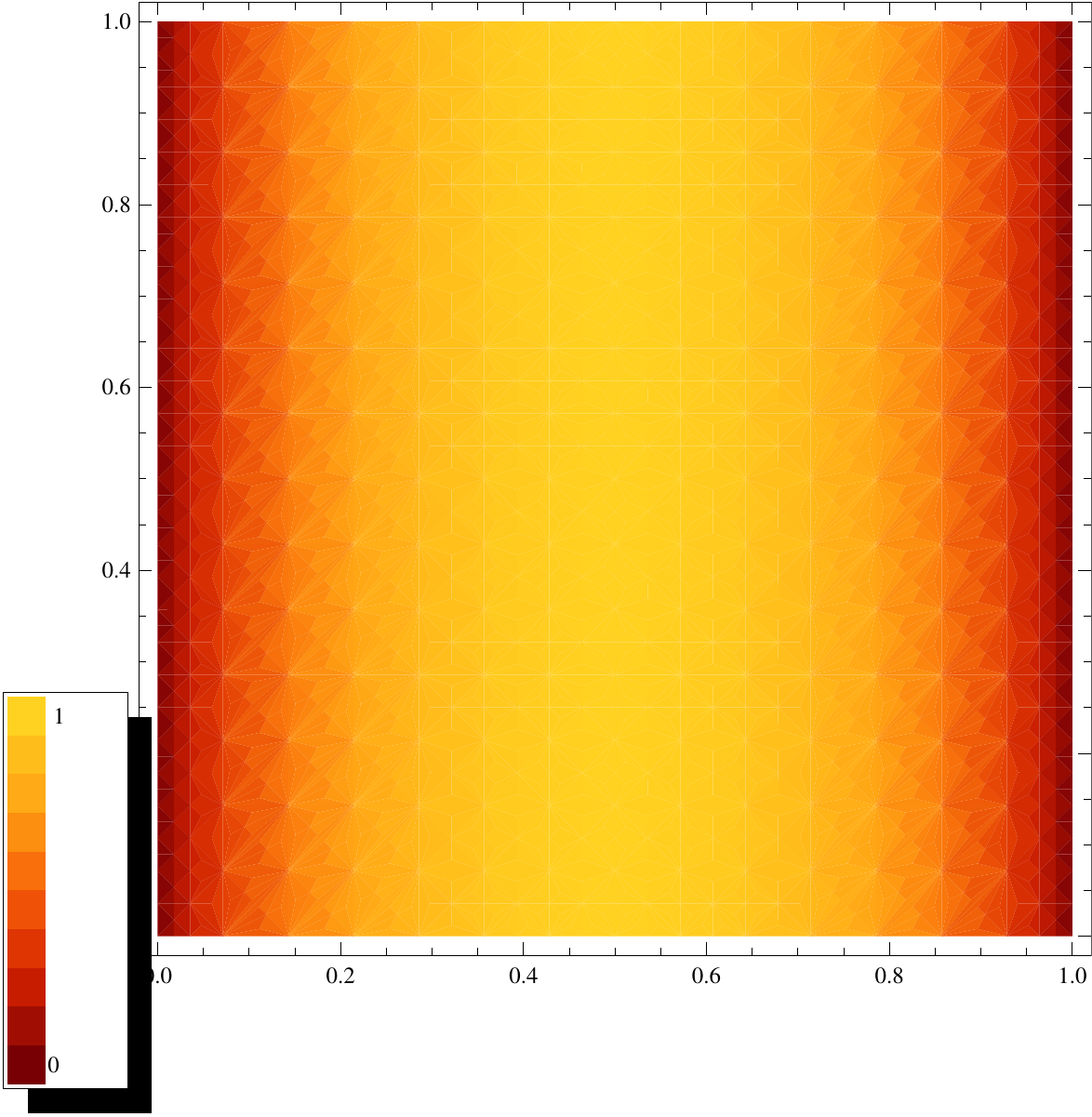}
\caption{Density plot of the Left figure.} \label{Cij1d}
\end{minipage}\end{figure}

  For case (i),  conditional probabilities
$\Pr(\beta|a_{i},b=i)$ $i=0,1$ are symmetric. Moreover, since
$\xi_0=1$ and $\xi_1=0$ so that the
corresponding channel between
$a_{0}$ and $\beta$ for $\xi_0$ is noiseless and the
corresponding channel between
 $a_{1}$ and $\beta$ for $\xi_1$ is completely noisy. This then
leads to $I_1=0$ and $I=I_0$.
   The dependence of $I=I_0$ on one of the input marginal probabilities, i.e., $\Pr(a_0=0)$ only, is shown in Fig \ref{Cij1I0}-\ref{Cij1d}.
  Note that $I$ reaches its maximal value, $1$ at $\Pr(a_0=0)=\frac{1}{2}$ as expected for the symmetric  conditional probabilities
$\Pr(\beta|a_{0},b=0)$ with $\xi_0=1$.
   This point is nothing but the point of maximal $I$ in Fig \ref{d2k2g}.
   Note that this maximum saturates the bound by information causality.
    This implies that we can reach the causally-allowed bound on  information gain by sacrificing one of  the sub-set of the conditional
    probabilities,
$\Pr(\beta|a_{1},b=1)$, without any comprise.
     This is a bit surprising.

\begin{figure}
\begin{minipage}[t]{0.5\linewidth}
\centering
\includegraphics[width=3in]{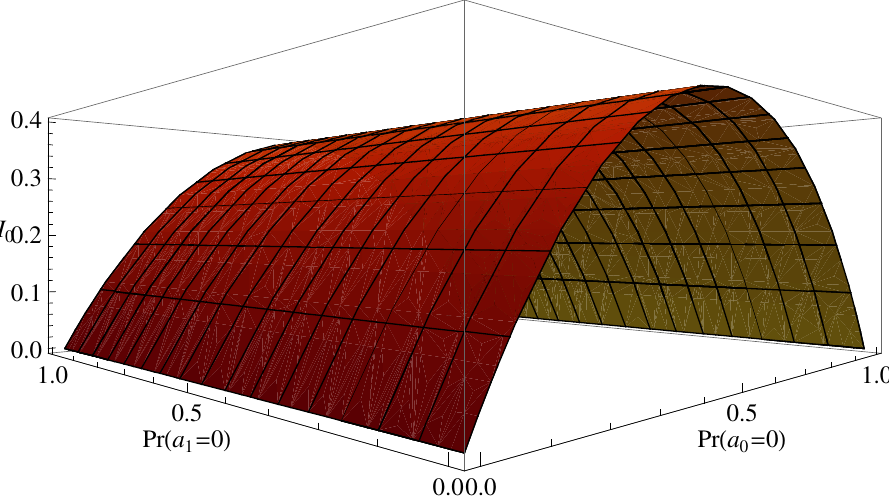}
\caption{$I_0$ vs $\Pr(a_{0,1}=0$) for case (ii).} \label{CijtI0}
\end{minipage}%
\begin{minipage}[t]{0.5\linewidth}
\centering
\includegraphics[width=3in]{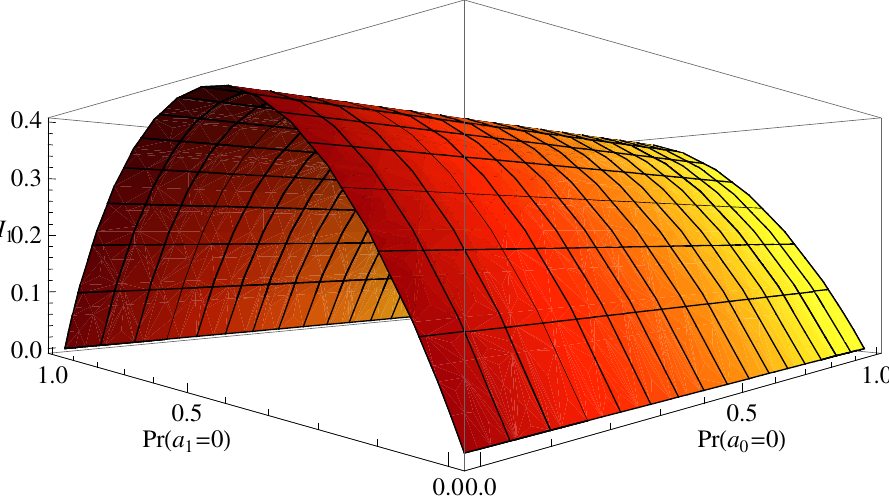}
\caption{$I_1$ vs $\Pr(a_{0,1}=0$) for case (ii).} \label{CijtI1}
\end{minipage}
\begin{minipage}[t]{0.5\linewidth}
\centering
\includegraphics[width=3in]{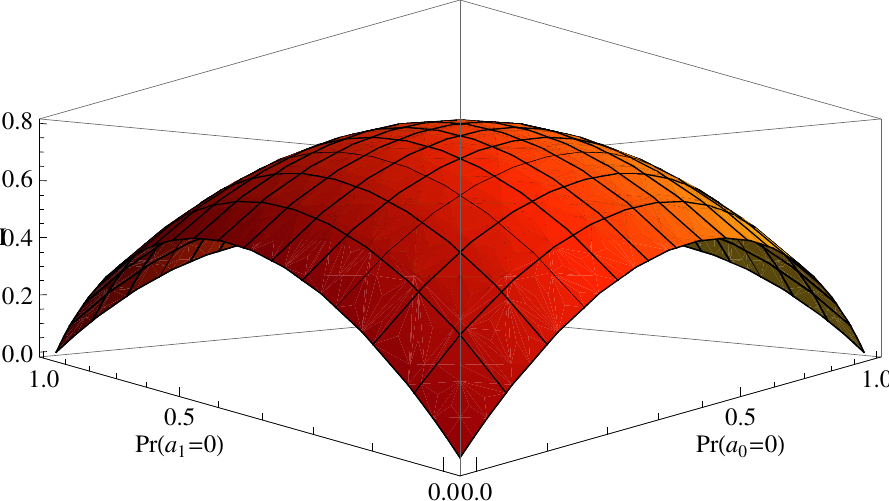}
\caption{$I$ vs $\Pr(a_{0,1}=0$) for case (ii).} \label{Cijt}
\end{minipage}%
\begin{minipage}[t]{0.5\linewidth}
\centering
\includegraphics[width=2.3in]{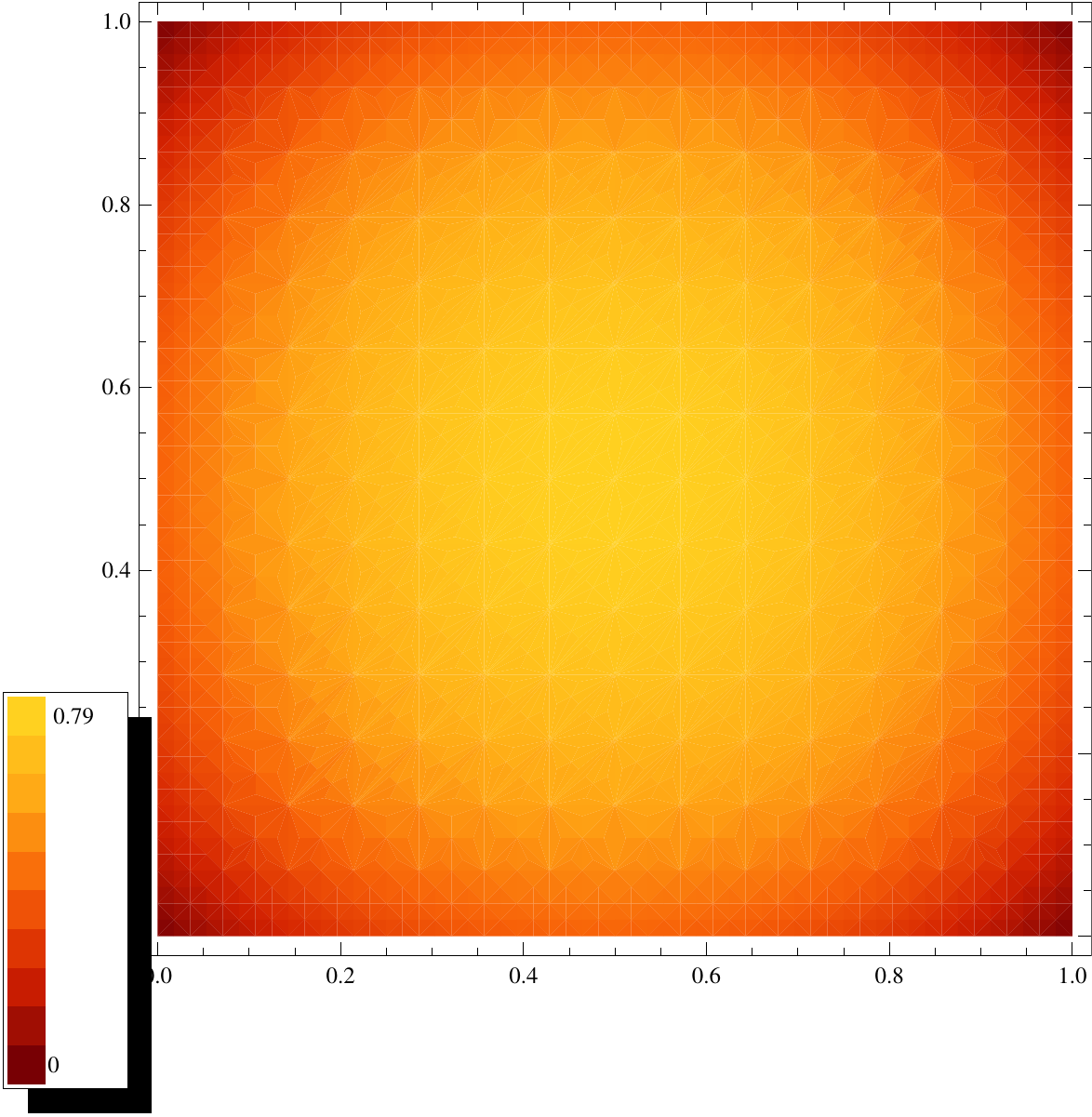}
\caption{Density plot of the Left figure.} \label{Cijtd}
\end{minipage}
\end{figure}

   For case (ii),  the conditional probabilities
$\Pr(\beta|a_{i},b=i)$ for $i=0,1$ are both symmetric and
isotropic, we then expect that the isotropy will also appear in the
plot for $I$ vs the input marginal probabilities $\Pr(a_{i})$, and
that $I_0$ and $I_1$ will have the same shape. This is indeed the
case as shown in Fig \ref{CijtI0}-\ref{Cijtd}. Note that $I_i$ only
depends on $\Pr(a_i)$ though $I=I_0+I_1$ depends on both. We see
that the maximal value of $I$ occurs at the symmetric point, i.e.,
all the $Pr(a_i)$ equal to $\frac{1}{2}$. However, the maximal value
is $0.7983$ which is less than $1$ of the information causality but
is the same value for the case of the Tsirelson bound.

\begin{figure}
\begin{minipage}[t]{0.5\linewidth}
\centering
\includegraphics[width=3in]{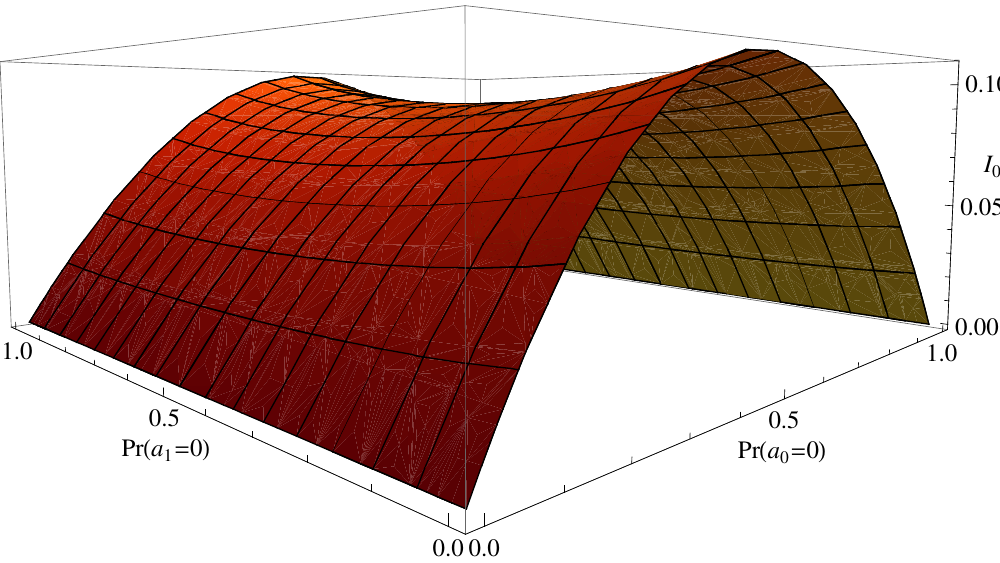}
\caption{$I_0$ vs $\Pr(a_{0,1}=0$) for case (iii).} \label{Cij4I0}
\end{minipage}%
\begin{minipage}[t]{0.5\linewidth}
\centering
\includegraphics[width=3in]{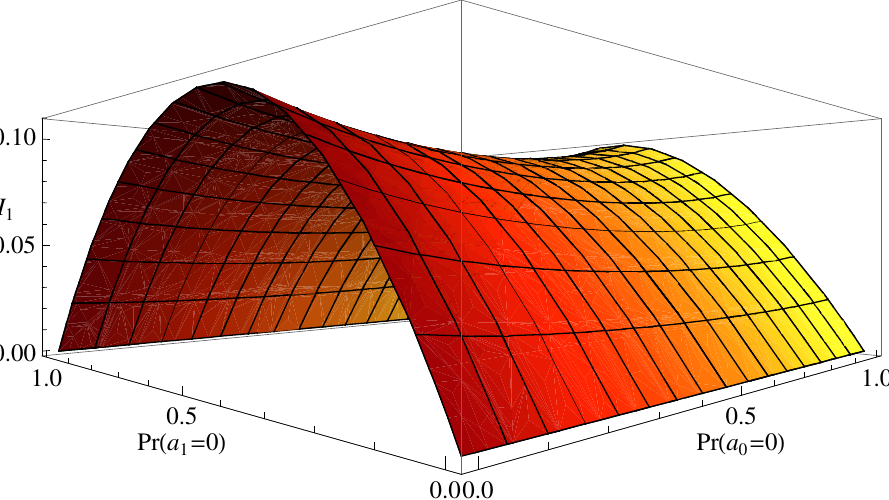}
\caption{$I_1$ vs $\Pr(a_{0,1}=0$) for case (iii).} \label{Cij4I1}
\end{minipage}
\begin{minipage}[t]{0.5\linewidth}
\centering
\includegraphics[width=3in]{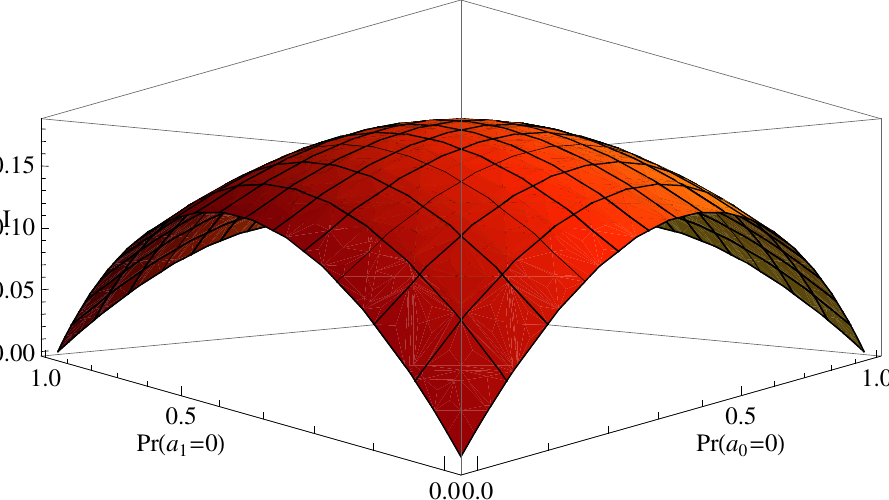}
\caption{$I$ vs $\Pr(a_{0,1}=0$) for case (iii).} \label{Cij4}
\end{minipage}%
\begin{minipage}[t]{0.5\linewidth}
\centering
\includegraphics[width=2.3in]{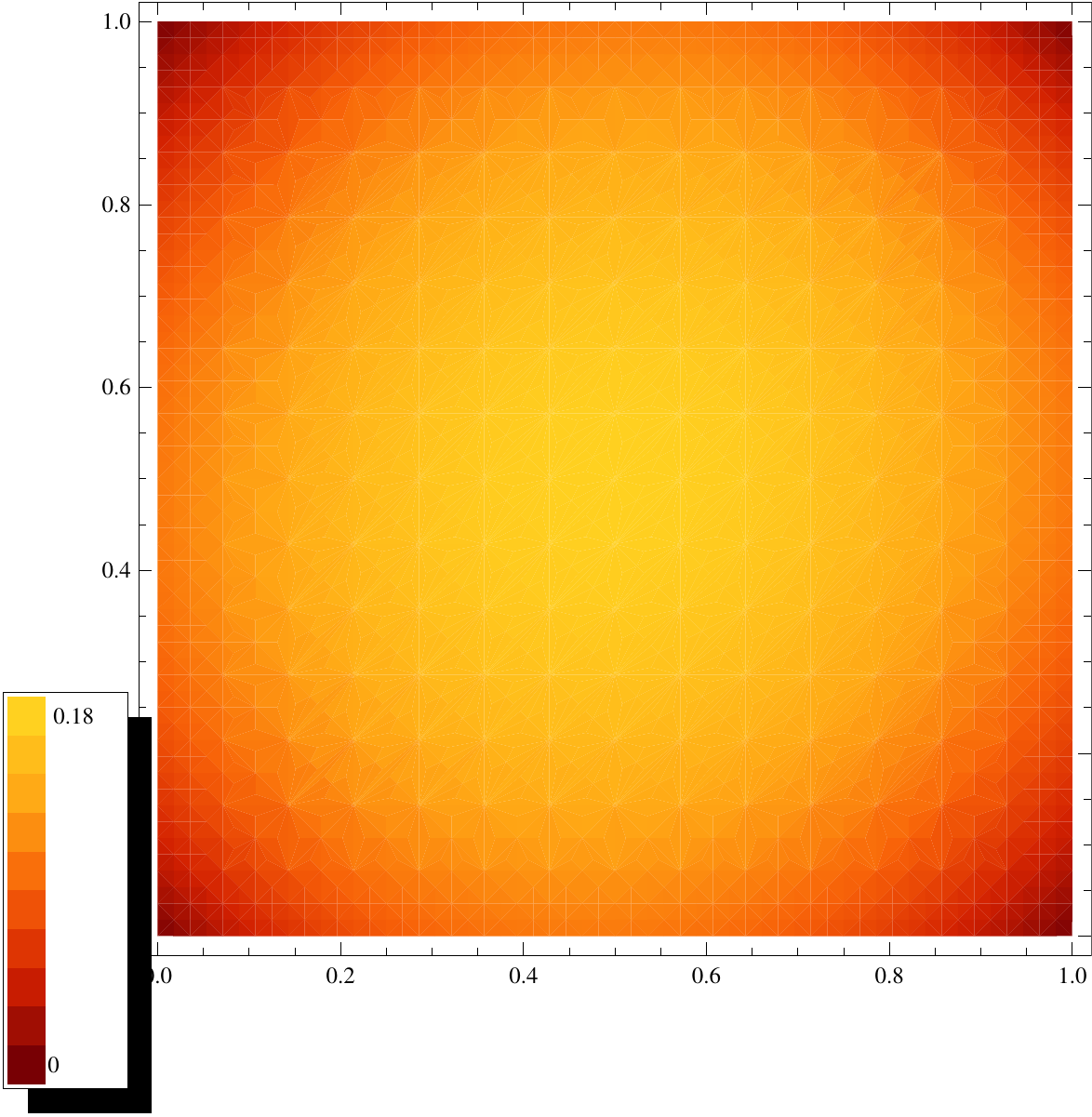}
\caption{Density plot of the Left figure.} \label{Cij4d}
\end{minipage}
\end{figure}

   Finally, for case (iii), i.e., the particular asymmetric  conditional probabilities
$\Pr(\beta|a_{i},b=i)$, $I_i$'s are now dependent on both
$\Pr(a_i)$'s unlike in the previous two cases.
    However, the  information gain $I$ has the isotropic form as in the case (ii) but with a far smaller maximal value at the symmetric point.
The results are shown in Fig \ref{Cij4I0}-\ref{Cij4d}.

   Our above results implies that the closer to $1$ is the $\Pr(B_{y}-A_{x}=xy|x,y)$, the larger is the information gain $I$. This is consistent with our RAC protocol as Bob can perfectly guess Alice's inputs by using the PR box \cite{PRbox}. Of course, the information causality ensures that the NS-box constrained by quantum mechanics can not be the PR box.
    Also, note that the maximum of $I$ occurs at the symmetric point of the input marginal probabilities $\Pr(a_{i})$ for case (ii) and (iii) but it is not the case for case (i). Therefore, the uniform input marginal probabilities $\Pr(a_{i})$ do not always lead to the maximal $I$.

\subsection{Information causality for the most general conditional probabilities}

   After testing the information causality for the more general conditional probabilities
$\Pr(\beta|a_{i},b=i)$ as discussed in the previous sections,
    we would wonder if the information causality holds for the most general conditional probabilities
$\Pr(\beta|a_{i},b=i)$ or not,
   i.e., $\Pr(\beta|a_{i},b=i)$ without any additional constraint on the
joint probabilities of the NS-box and the input marginal
probabilities $\Pr(a_{i})$ except the necessary quantum and
no-signaling constraints.
      For our $d=2$, $k=2$ RAC protocol,
       we check this by partitioning the defining domains of the probabilities into 100 points and then using the brute-force method to do the numerical check.
        We find that the information causality is always satisfied. This yields a more general support for the information causality.

    Furthermore,
     we find that the information causality is saturated, i.e., $I=1$ when one of  the sub-sets of conditional probabilities
$\Pr(\beta|a_{i},b=i)$ corresponds to the noiseless channel between
$a_{i}$ and $\beta$ and the other one corresponds to completely
noisy channel.
      This is similar to the case (i) discussed in the previous subsection.

\section{Conclusion}\label{sec6}

   Information causality was proposed as a new physical principle and gives an intuitive picture on the meaning of causality from the information point of view.
    Therefore, to test its validity for general communication schemes will help to establish it as a physical principle.
     Motivated by this, in this work we try our best to extend the framework of the original proposal to the more general cases, such as the multi-level and multi-setting RAC protocols or lifting the  symmetric and isotropic conditions on the conditional probabilities
$\Pr(\beta|a_{i},b=i)$ or uniform condition on the  input marginal probabilities
$\Pr(a_{i})$.
       We then test the information causality for these general protocols by either adopting the SDP for numerical check,
        or using the brutal force method for the more general  conditional probabilities
$\Pr(\beta|a_{i},b=i)$.
         With all these efforts, our results are rewarding: we see that the information causality are preserved in all the protocols discussed in this work.
           This reinforce the validity of the information causality further than before.
            Though more checks for more general protocols should be always welcome.
             We also find that the information causality is saturated not by sharing the correlations saturating the Tsireslon bound, but by the ones which saturate the CHSH (or Bell) inequality.
              This then raises the issues on the intimate relation between the information gain and the quantum violation of the Bell-type inequalities.
               Especially, this result challenges our intuition that a channel can transfer more information by the quantum resources with the more amount of the violation of the Bell-type inequalities.
                We think our findings in this paper will shed some light on the related topics.

\begin{acknowledgements}
This project is supported by Taiwan's NSC grants (grant NO.
100-2811-M-003-011 and 100-2918-I-003-008).

\end{acknowledgements}


\begin{appendix}
\section{Signal decay and data processing inequality for multi-nary channels}\label{app-1}

   In this appendix, we will first sketch the key steps of \cite{Evans1} in obtaining the maximal bound on the signal decay for the binary noisy channels, and then generalize this derivation to the one for the multi-nary channels.

   Our setup is to consider a cascade of two communication channels: $X\rightarrow
Y\rightarrow Z$. The decay of the signal is implied by the  data
processing inequality, i.e., \be\label{DP} I(X;Z)\leq I(X;Y). \ee
The mutual information $I(X;Y)=H(Y)-\sum_i \Pr(X=i) H(Y|X=i)$, where
$H(Y)$ and $H(Y|X)$ are the Shannon entropies for the probabilities
$\Pr(Y)$ and the conditional probabilities $\Pr(Y|X)$, respectively.

Furthermore, for the binary symmetric channel $A$ characterized by
\bea\label{binaryA} A= \left ( {\begin{array}{cc}
      \frac{1+\xi}{2}&\frac{1-\xi}{2}     \\
      \frac{1-\xi}{2}&\frac{1+\xi}{2}      \\
      \end{array} } \right),
\eea it was shown in \cite{Evans1} that the bound on the signal
decay is characterized by the following bound \be\label{bBound}
\frac{I(X;Z)}{I(X;Y)}\le \xi^2. \ee Note that this bound is tighter
than the one obtained in \cite{Pi}, which is
$\frac{I(X;Z)}{I(X;Y)}\le \xi$.

    In this appendix, we will generalize the above result to the one for the dinary channel characterized by $\Pr(Z=i|Y=i)=\xi$ and $\Pr(Z=s\ne i|Y=i)=\frac{1-\xi}{d-1}$ with $i\in\{0,1,...,d-1\}$,  so that the signal decay is bound by
\be\label{dST4b} \frac{I(X;Z)}{I(X;Y)}\leq(\xi-\frac{1-\xi}{d-1})^2.
\ee

\subsection{Sketch of the proof in \cite{Evans1}}

     The derivation in \cite{Evans1} consists of two key steps. The first one is to show the following theorem for weak signal:

{\it Theorem I: The ratio $\frac{I(X;Z)}{I(X;Y)}$ reaches its
maximum if the conditional probabilities $\Pr(Y|X=0)$ and
$\Pr(Y|X=1)$ are almost indistinguishable, i.e.,
$|\Pr(Y=0|X=0)-\Pr(Y=0|X=1)|\rightarrow0$. }

\bigskip

   To prove this theorem we need the following lemma:

{\it Lemma I: For any strictly concave function $f$ and $g$ on the
interval $[0,1]$, and any $p\in [0,1]$, the ratio \be
r(x,y)=g_2(x,y,p)/f_2(x,y,p) \ee reaches its maximum in the limit
$|x-y|\rightarrow 0$. Here $f_2(x,y,p)=f(px+(1-p)y)-pf(x)-(1-p)f(y)$
denotes the second order difference of the function $f$ with the
weight $p$, and similarly for the $g_2(x,y,p)$.}

 \bigskip

   We sketch the proof of this lemma, which will be useful when generalizing to the multi-nary channel.  We assume that the ratio $r$ reaches its maximum at $x=x^*$ and $y=y^*$, and for concreteness assuming $x^*<y^*$. Note that $0<r<\infty$ due to the concave $f$ and $g$. We can perform affine transformation to scale this maximal value of $r(x^*,y^*,p)$ to be $1$, and also to make $f(x^*)=g(x^*)$ and $f(y^*)=g(y^*)$. This immediately leads to $f(px^*+(1-p)y^*)=g(px^*+(1-p)y^*)$. That is, there is a point $z^*=px^*+(1-p)y^*$ inside the interval $[x^*,y^*]$ at which $f$ also equals to $g$.   Use this fact, it is easy to convince oneself that either $r(z^*,y^*)\ge r(x^*,y^*)$ or $r(x^*,z^*)\ge r(x^*,y^*)$.  For more subtle details, please see \cite{Evans1}. By repeating this procedure we prove the lemma.

  Observe that $I(X;Y)$ and $I(X;Z)$ are the second order difference of the (concave) entropy functions $H(Y)$ and $H(Z)$, respectively with the weight $p=\Pr(X=0)$. We can then prove the Theorem I by the above lemma.

\bigskip

   The second step is first to rewrite the ratio $\frac{I(X;Z)}{I(X;Y)}$ in terms of relative entropy  $D(p\|q):=\sum_{x}\Pr(p=x)log\frac{\Pr(p=x)}{\Pr(q=x)}$,
    that is,
\begin{eqnarray}\label{rID}
\frac{I(X;Z)}{I(X;Y)}=\frac{\sum_{i=0}^{1}\Pr(X=i)D(\Pr(Y|X=i)\cdot
A\|\Pr(Y)\cdot A)}{\sum_{i=0}^{1}\Pr(X=i)D(\Pr(Y|X=i)\|\Pr(Y))}.
\end{eqnarray}
Then, based on the above theorem we can parameterize the conditional
probabilities $\Pr(Y|X=0)=\vec{p}+\vec{\epsilon}$ where
$\vec{p}=\sum_{i=0}^{1}\Pr(X=i)\Pr(Y|X=i)$   and
$\vec{\epsilon}=(\epsilon,-\epsilon)$ with $\epsilon$ being
sufficiently small. With this condition, \eq{rID} can be simplified
to
\begin{eqnarray}\label{ST2}
\frac{I(X;Z)}{I(X;Y)}\approx\frac{D((\vec{p}+\vec{\epsilon})\cdot
A\parallel \vec{p}\cdot
A)}{D(\vec{p}+\vec{\epsilon}\parallel\vec{p})}.
\end{eqnarray}
Note that the ratio now does not depend on $\Pr(X)$.

   Finally, given the binary channel \eq{binaryA} we can expand the relative entropy in terms of $\epsilon/\Pr(Y)$, so for the ratio $\frac{I(X;Z)}{I(X;Y)}$. Then, fixing $\epsilon$ and then varying the first order term of the ratio $\frac{I(X;Z)}{I(X;Y)}$ in the above expansion over  $\vec{p}$, we obtain the bound in \eq{bBound}.

 \subsection{Generalizing to the multi-nary channels}

    We now generalize the above derivation to the trinary noisy channels, then the generalization to the dinary channel will just follows. The key steps are similar to the binary ones. The first step is to use the same method to prove the following theorem:

{\it  Theorem II: The ratio $\frac{I(X;Z)}{I(X;Y)}$ reaches its
maximum only when all the three conditional probabilities
$\Pr(Y|X=i)$ with $i=0,1,2$ are  almost indistinguishable. }

\bigskip

  The strategy to prove this theorem is to observe that we can treat the pair $(\Pr(Y=0|X=i),\Pr(Y=1|X=i))$ for each $i$ (note that $\Pr(Y=2|X=i)$ is not independent of this pair) as a point inside the unit square $([0,1],[0,1])$. Then the three points $\Pr(Y|X=i)$ for $i=0,1,2$ form a triangle. We can then follow the same way of proving the Lemma I in the previous subsection for the trinary case.  First, we assume the maximal value of $r$ occurs at all three vertices of some triangle. We then perform the affine transformation to rescale this maximal value to $1$, and to make  $f=g$ (or more specifically $H(Y|X=i)=H(Z|X=i)$) at the three vertices of the above triangle.  This then immediately leads to that there exists some point inside the triangle such that $f=g$.
    We can use this point to construct a smaller triangle with any two of the vertices of the original triangle and show that the ratio $r$ for this new triangle is greater than the one for the original larger triangle.  Repeating this procedure we can prove the above theorem. It is also clear that we can generalize the theorem for the multi-nary channels by generalizing the triangle to the concave body of the higher dimensional space.

  Here, we should point out that one can always reduce the concave body to the linear interval one, so that we can reduce to the situation for the binary case. That is, we set all the conditional probabilities except one to be equal, and then study the closeness condition of the remaining two distinct conditional probabilities for the maximal ratio of $\frac{I(X;Z)}{I(X;Y)}$. In the following, we will always restrict to such a situation.

\bigskip

    We then go to the second step as for the binary channel, that is to use Theorem II to reduce the problem of maximizing $\frac{I(X;Z)}{I(X;Y)}$ to the one of  maximizing the ratio of relative
entropies. We rewrite the ratio of two mutual information as
following,
\begin{eqnarray}\label{rIID}
&\frac{I(X;Z)}{I(X;Y)}=\frac{\sum_{i=0}^{2}\Pr(X=i)D(\Pr(Y|X=i)\cdot
A\|\Pr(Y)\cdot A)}{\sum_{i=0}^{2}\Pr(X=i)D(\Pr(Y|X=i)\|\Pr(Y))}.
\end{eqnarray}

 To simplify the expression for further manipulations, we denote the average probability distribution of $Y$ as $\vec{p}=\sum_{i=0}^{2}\Pr(X=i)\Pr(Y|X=i)$, and parameterize the probability $\Pr(Y|X=0)=\vec{p}+\vec{\epsilon}_{0}$ and $\Pr(Y|X=1)=\vec{p}+\vec{\epsilon}_{1}$. Thus, the probability $\Pr(Y|X=2)$ is forced to be $\vec{p}-\frac{\Pr(X=0)}{\Pr(X=2)}\vec{\epsilon}_{0}-\frac{\Pr(X=1)}{\Pr(X=2)}\vec{\epsilon}_{1}$. The parameter vectors $\vec{\epsilon}_0$ and $\vec{\epsilon}_1$ should be sufficiently small as required by Theorem II to have maximal ratio $\frac{I(X;Z)}{I(X;Y)}$.  Furthermore, we will further reduce the triangle to the linear interval case by assuming $\vec{\epsilon}_0=\vec{\epsilon}_1$, i.e., $\Pr(Y|X=0)=\Pr(Y|X=1)$.

The ratio \eq{rIID} then becomes
\begin{eqnarray}\label{ST2}
\frac{I(X;Z)}{I(X;Y)}\approx\frac{D((\vec{p}+\vec{\epsilon_{0}})\cdot
A\parallel \vec{p}\cdot
A)}{D(\vec{p}+\vec{\epsilon_{0}}\parallel\vec{p})}.
\end{eqnarray}
Note again the ratio now does not depend on $\Pr(X)$.

Before serious expansion of \eq{ST2} in the power of
$\vec{\epsilon}_0$, we need to specify
$\vec{p}=({\scriptstyle\Pr(Y=0),\Pr(Y=1),\Pr(Y=2)})$ and
$\vec{\epsilon}_{0}=({\scriptstyle v_{0},v_{1},v_{2}})$. Note that,
$v_{0}+v_{1}+v_{2}=0$. As for the bi-nary channel, we expand the
relative entropy in terms of $\frac{v_{i}}{\Pr(Y=i)}$. The leading
term of the expansion for the denominator of \eq{ST2} is found to be
\begin{eqnarray}\label{ST3a}
D(\vec{p}+\epsilon_{0}\parallel\vec{p})=\frac{1}{2ln2}\sum_{i=0}^{1}\frac{v_{i}^2}{\Pr(Y=i)}.
\end{eqnarray}

To find the expansion of the numerator, we need to specify the
channel $A$ between $Y$ and $Z$.  The generic trinary channel is
given by
\begin{eqnarray}
A=\Pr(Z|Y)=    \left ( {\begin{array}{ccc}
      a_{1}&a_{2}&a_{3}     \\
      b_{1}&b_{2}&b_{3}      \\
      c_{1}&c_{2}&c_{3}
      \end{array} } \right),
\end{eqnarray}
where the elements of the channel should satisfy
$a_{1}+a_{2}+a_{3}=1$, $b_{1}+b_{2}+b_{3}=1$, and
$c_{1}+c_{2}+c_{3}=1$. Then, the leading term in the expansion of
the numerator of \eq{ST2} is found to be
\begin{eqnarray}\label{ST3b}
{\textstyle D((\vec{p}+\vec{\epsilon_{0}})\cdot A\parallel
\vec{p}\cdot
A)=\frac{1}{2ln2}(\frac{v_{0}a_{1}+v_{1}b_{1}+v_{2}c_{1}}{p(Z=0)}+\frac{v_{0}a_{2}+v_{1}b_{2}+v_{2}c_{2}}{p(Z=1)}+\frac{v_{0}a_{3}+v_{1}b_{3}+v_{2}c_{3}}{p(Z=2)})}.
\end{eqnarray}
For simplicity, we only consider the symmetry trinary channel as
follows
\begin{eqnarray}
A=\Pr(Z|Y)=
      \left ( {\begin{array}{ccc}
      \xi&\frac{1-\xi}{2}&\frac{1-\xi}{2}     \\
      \frac{1-\xi}{2}&\xi&\frac{1-\xi}{2}      \\
      \frac{1-\xi}{2}&\frac{1-\xi}{2}&\xi
      \end{array} } \right).
\end{eqnarray}
Then, \eq{ST3b} then becomes
\begin{eqnarray}\label{ST4a}
D((\vec{p}+\vec{\epsilon_{0}})\cdot A\parallel \vec{p}\cdot
A)=(\frac{3\xi-1}{2})^2\frac{1}{2ln2}\sum_{i=0}^{2}\frac{v_{i}^2}{\Pr(Z=i)}.
\end{eqnarray}

Since we know that for symmetric channel, the maximal mutual
information is achieved for uniform input probabilities. Thus, we
assume uniform $\Pr(Y)$ and $\Pr(Z)$ so that \eq{ST2} depends only
on variable $\xi$. We then obtain
\begin{eqnarray}\label{ST4b}
\frac{I(X;Z)}{I(X;Y)}\leq(\frac{3\xi-1}{2})^2.
\end{eqnarray}
This is the generalization of \eq{bBound} for binary channel to the
trinary one.

  Similarly, we can generalize the above derivation to the dinary channels.  If the channel between $Y$ and $Z$ is a dinary and symmetry channel specified as follows: $\Pr(Z=i|Y=i)=\xi$ and $\Pr(Z=s\ne i|Y=i)=\frac{1-\xi}{d-1}$ with $i\in\{0,1,...,d-1\}$, then the bound of the ratio $\frac{I(X;Z)}{I(X;Y)}$ is given by \eq{dST4b}.

\section{The concavity of information gain}
In this appendix, we want to prove the   information gain $I$ is not
a concave function to joint probabilities
$\Pr(B_{\vec{y}}-A_{\vec{x}}|\vec{x},\vec{y})$ and input marginal
probabilities $\Pr(a_{i})$. Thus, we could not formulate the problem
(maximizing  information gain $I$) as a convex optimization
programming.

First, we re-express  information gain $I$ by
$\Pr(B_{\vec{y}}-A_{\vec{x}}|\vec{x},\vec{y})$ and $\Pr(a_{i})$. If
maximizing information gain is a concave function to these
probabilities, the second order partial derivative of mutual
information respecting to each probability should be negative. Here,
we find a violation when calculating $\frac{\partial^{2}
I}{\partial(\Pr(B_{\vec{y}}-A_{\vec{x}}=0|\vec{x}=0,\vec{y}=0))^2}$.
In following paragraphs, we denote the joint probabilities
$\Pr(B_{\vec{y}}-A_{\vec{x}}=0|\vec{x}=0,\vec{y}=0)$ as $V$.

The   information gain can be rewritten as
\begin{eqnarray}
I=\sum_{i=0}^{k-1}I_{b=i},
\end{eqnarray}
where $I_{b=i}$ is equal to $I(a_{i};\beta|b=i)$. Since the joint
probability $V$ only contribute to $I_{b=0}$, we only need to
calculate $\frac{\partial^{2} I_{b=0}}{\partial V^2}$. The
reexpression of $I_{b=0}$ is
\begin{eqnarray}
I_{b=0}=\sum_{n=0}^{d-1}\sum_{j=0}^{d-1}\Pr(\beta=n,a_{0}=j|b=0)log_{2}\frac{\Pr(\beta=n,a_{0}=j|b=0)}{\Pr(\beta=n|b=0)\Pr(a_{0}=j|b=0)}.
\end{eqnarray}
Therefore, the first order partial derivative respecting to
$\Pr(B_{\vec{y}}-A_{\vec{x}}=0|\vec{x}=0,\vec{y}=0)$ is
\begin{eqnarray}\label{1stpd}
&&\frac{\partial I_{b=0}}{\partial V}=\notag\\
 &&\sum_{n=0}^{d-1}\sum_{j=0}^{d-1}\frac{\partial
\Pr(a_{0}=j,\beta=n|b=0)}{\partial V}log_{2}\frac{\Pr(a_{0}=j,\beta=n|b=0)}{\Pr(\beta=n|b=0)\Pr(a_{0}=j|b=0)}\notag\\
 &&+\frac{1}{ln2}(\frac{\partial
\Pr(a_{0}=j,\beta=n|b=0)}{\partial
V}-\frac{\Pr(a_{0}=j,\beta=n|b=0)}{\Pr(\beta=n|b=0)}\frac{\partial
\Pr(\beta=n|b=0)}{\partial V})
\end{eqnarray}

We can express $\Pr(a_{0}=j,\beta=n|b=0)$ as the combination of
joint probabilities $\Pr(B_{\vec{y}}-A_{\vec{x}}|\vec{x},\vec{y})$
and input marginal probabilities $\Pr(a_{i})$ to obtain
$\frac{\partial
\Pr(a_{0}=j,\beta=n|b=0)}{\partial V}$.\\
Since joint probabilities $
\Pr(B_{\vec{y}}-A_{\vec{x}}|\vec{x},\vec{y})$ are subjected to the
normalization conditions of total probability,
 if $n-j\neq(d-1)$,
\begin{equation}
{\scriptstyle\Pr(a_{0}=j,\beta=n|b=0)=\sum_{a_{k\neq0}}\Pr(B_{\vec{y}}-A_{\vec{x}}=n-j|\vec{x},\vec{y}=0)\Pr(a_{0}=j)\;
\Pi_{k\ne 0} \Pr(a_k)};
\end{equation}
 if $n-j=(d-1)$,
\begin{equation}
{\scriptstyle\Pr(a_{0}=j,\beta=n|b=0)=\sum_{a_{k\neq0}}(1-\sum_{t=0}^{d-2}\Pr(B_{\vec{y}}-A_{\vec{x}}=t|\vec{x},\vec{y}=0))\Pr(a_{0}=j)\;
\Pi_{k\ne 0} \Pr(a_k)},
\end{equation}
where $\vec{x}$ in the above functions is given by the RAC encoding, i.e.,  $\vec{x}:=(x_{1},\cdots,x_{k-1})$ with $x_{i}=a_{i}-a_{0}$

Now, we can calculate the derivatives. The patrial derivative
\begin{eqnarray}\label{1std}
\frac{\partial \Pr(a_{0}=j,\beta=n|b=0)}{\partial V}
\end{eqnarray}
is not equal to zero for two cases, the first one is $j=n$, we can
obtain $\; \Pi_{k} \Pr(a_k=n)$ for \eq{1std}. The second case is
$n-j=(d-1)$, we can obtain $\; -\Pi_{k} \Pr(a_k=n-(d-1))$.
Therefore, since
$\Pr(\beta=n|b=0)=\sum_{j}\Pr(a_{0}=j,\beta=n|b=0)$, we can obtain
\begin{eqnarray}
\frac{\partial \Pr(\beta=n|b=0)}{\partial V}&&=\Pi_{k}
\Pr(a_k=n)-\Pi_{k} \Pr(a_k=n-(d-1)).
\end{eqnarray}
Put above result to \eq{1stpd}, for fixed $j$, we can find that
$\sum_{n=0}^{d-1}\frac{\partial \Pr(a_{0}=j,\beta=n|b=0)}{\partial
V}=0$, thus the second term of \eq{1stpd} will vanish.

We then can calculate the second order derivative
\begin{eqnarray}\label{2end}
&\frac{\partial^{2} I_{b=0}}{\partial
V^{2}}=\frac{1}{ln2}\sum_{n=0}^{d-1}\sum_{j=0}^{d-1}(\frac{\partial
\Pr(a_{0}=j,\beta=n|b=0)}{\partial V})^{2}\frac{1}{\Pr(a_{0}=j,\beta=n|b=0)}\notag\\
 &-\frac{2}{\Pr(\beta=n|b=0)}\frac{\partial
\Pr(a_{0}=j,\beta=n|b=0)}{\partial V}\frac{\partial
\Pr(\beta=n|b=0)}{\partial V}+(\frac{\partial
\Pr(\beta=n|b=0)}{\partial
V})^2\frac{\Pr(a_{0}=j,\beta=n|b=0)}{(\Pr(\beta=n|b=0))^2}
\end{eqnarray}

For $d=2$ and $k=2$, \eq{2end} becomes
\begin{eqnarray}
&&\frac{\partial^{2}
I}{\partial V^{2}}=\frac{1}{ln2}[(\Pr(a_{0}=0)\Pr(a_{1}=0))^{2}(\frac{1}{\Pr(a_{0}=0,\beta=0|b=0)}+\frac{1}{\Pr(a_{0}=0,\beta=1|b=0)})\notag\\
&&+(\Pr(a_{0}=1)\Pr(a_{1}=1))^{2}(\frac{1}{\Pr(a_{0}=1,\beta=0|b=0)}+\frac{1}{\Pr(a_{0}=1,\beta=1|b=0)})\notag\\
&&-(\frac{1}{\Pr(\beta=0|b=0)}+\frac{1}{\Pr(\beta=1|b=0)})(\Pr(a_{0}=0)\Pr(a_{1}=0)-\Pr(a_{0}=1)\Pr(a_{1}=1))^{2}]\notag\\
\end{eqnarray}
Once $\Pr(a_{0}=0)=1-\Pr(a_{1}=0)$, the above function is
non-negative.

For higher $d$ and $k$, once the input marginal probabilities
$\Pr(a_{i})$ are uniform. We then can obtain
\begin{eqnarray}
&&\frac{\partial^{2} I}{\partial V^{2}}=\frac{\partial^{2} I_{b=0}}{\partial V^{2}}=\notag\\
 &&\frac{1}{ln2}\sum_{n=0}^{d-1}\frac{1}{d^{2k}}(\frac{1}{\Pr(a_{0}=n,\beta=n|b=0)}+\frac{1}{\Pr(a_{0}=n,\beta=n-(d-1)|b=0)})\notag\\
 &&>0
\end{eqnarray}

It is clear that  information gain $I$ is not a concave function to
joint probabilities $\Pr(B_{\vec{y}}-A_{\vec{x}}|\vec{x},\vec{y})$
and input marginal probabilities $\Pr(a_{i})$.

\section{Semidefinite programming}\label{sdp}

In this appendix, we briefly introduce the semidefinite programming
(SDP) \cite{SDP}. SDP is the problem of optimizing a linear function
subjected to certain conditions associated with a positive
semidefinite matrix $X$, i.e., $v^{\dag}Xv\geq0$, for $v
\in\mathbb{C}^{n}$, and is denoted by $X\succeq0$. It can be
formulated as the standard primal problem as follows. Given the
$n\times n$ symmetric matrices $C$ and $D_{q}$'s with
$q=1,\cdots,m$, we like to optimize the $n\times n$ positive
semidefinite matrix $X\succeq0$ such that we can achieve the
following:
\begin{subequations}
\label{cons}%
\begin{align}
minimize \qquad &  \Tr(C^{T}X)\\
subject \quad to \qquad &  \Tr(D_{q}^{T}X)=b_{q}, \quad
q=1,\cdots,m\;.
\end{align}
\end{subequations}
Corresponding to the above primal problem, we can obtain a dual
problem via a Lagrange approach \cite{CV}. The Lagrange duality can
be understood as the following. If the primal problem is

\begin{subequations}
\begin{align}
minimize\qquad &  f_{0}(x)\\
s.t.\qquad     &  f_{q}(x)\leq0,\quad q\in1...m.\\
               &  h_{q}(x)=0,\quad q\in1...p,
\end{align}
\end{subequations}
the Lagrange function can be defined as%
\begin{equation}
L(x,\lambda,\nu)=f_{0}(x)+\Sigma_{q=1}^{m}\lambda_{q}f_{q}(x)+\Sigma_{q=1}%
^{p}\nu_{q}h_{q}(x), \label{Lf}%
\end{equation}
where $\lambda_{1}$,\ldots, $\lambda_{m}$, and
$\nu_{1}$,\ldots,$\nu_{p}$ are Lagrange multipliers respectively.
Due to the problem and (\ref{Lf}), the minima of $f_{0}$ is bounded
by (\ref{Lf}) under the constraints when $\lambda_{1}$,\ldots,
$\lambda_{m}\geq 0$.
\[
\mathop{\inf}_{x}f_{0}\geq\mathop{\inf}_{x}L(x,\lambda,\nu).
\]
Then the Lagrange dual function is obtained.
\[
g(\lambda,\nu)=\mathop{\inf}_{x}L(x,\lambda,\nu).
\]
$g(\lambda,\nu)\leq p$ ($p$ is the optimal solution of $f_{0}(x)$ ),
for $\lambda_{1}$,\ldots, $\lambda_{m}\geq0$ and arbitrary
$\nu_{1}$,\ldots ,$\nu_{p}$. The dual problem is defined.
\begin{subequations}
\label{Ldp}%
\begin{align}
maximize\qquad &  g(\lambda,\nu)\\
s.t.\qquad &  \lambda_{q}\geq0.\quad(q\in\{1...m\})
\end{align}
\end{subequations}
We can use the same method to define the dual problem for SDP. From
the primal problem of SDP (\ref{cons}), we can write down the dual
function by using
minimax inequality \cite{DPP}.%
\begin{widetext}
\begin{align}
\mathop{\inf}_{X\succeq0}\Tr(C^{T}X)  &  =\mathop{\inf}_{X\succeq
0}\Tr(C^{T}X)+\sum_{q=1}^{m}y_{q}(b_{q}-\Tr(D_{q}^{T}%
X))\nonumber\\
&  =\mathop{\inf}_{X\succeq0}\mathop{\sup}_{y}\sum_{q=1}^{m}y_{q}%
(b_{q})+\Tr((C^{T}-\sum_{q=1}^{m}y_{q}D_{q}^{T})X)\nonumber\\
&  \geq\mathop{\sup}_{y}\mathop{\inf}_{X\succeq0}\sum_{q=1}^{m}y_{q}%
(b_{q})+\Tr((C^{T}-\sum_{q=1}^{m}y_{q}D_{q}^{T})X)\nonumber\\
&  =\mathop{\sup}_{y}\mathop{\inf}_{X\succeq0}\sum_{q=1}^{m}y_{q}%
(b_{q})+\Tr((C-\sum_{q=1}^{m}y_{q}D_{q})^{T}X).
\end{align}
The optimal solution of dual function is bounded under some vector
$y$.
\begin{eqnarray}
\mathop{\sup}_{y} \mathop{\inf}_{X\succeq0}\sum_{q=1}^{m}
y_{q}(b_{q})+ \Tr((C-\sum_{q=1}^{m}y_{q}D_{q})^{T}X)= \left\{
                                                        \begin{array}{l}
                                                          \mathop{\sup}_{y}%
\sum_{q=1}^{m} y_{q}(b_{q}) \quad  ;when \quad
C-\sum_{q=1}^{m}y_{q}D_{q}
\succeq0 \quad \notag\\
                                                          -\infty\qquad ; otherwise.
                                                        \end{array}
                                                      \right.
 \end{eqnarray}
\end{widetext}
The correspond dual problem is
\begin{subequations}
\begin{align}
maximize \qquad &  \sum_{q=1}^{m} y_{q}(b_{q})\\
s.t. \qquad &  S=C-\sum_{q=1}^{m}y_{q}D_{q} \succeq0.
\end{align}
\end{subequations}

If the feasible solutions for the primal problem and the dual
problem attain their minimal and maximal values denoted as
$p^{\prime}$ and $d^{\prime}$ respectively, then $p^{\prime}\geq
d^{\prime}$, which is called the duality gap. This implies that the
optimal solution of primal problem is bounded by dual problem. This
then leads to the following: Both the primal and the dual problems
attain their optimal solutions when the duality gap vanishes, i.e.,
$d^{\prime}=p^{\prime}$.

\section{The quantum constraints for $n=1$ and $n=1+AB$
certificate} \label{higherSDP}

We divide this appendix into two parts. In the first part, we will
write down the associated quantum constraints for $\Gamma^{(1)}$ and
$\Gamma^{(1+AB)}$ when finding the bound of the Bell-type functions.
In the second part, we will estimate the number of these constraints
and find a efficient way to write down these constraints.
\subsection{The quantum constraints for $n=1$ and $n=1+AB$
certificate}

 When maximizing the Bell-type inequalities under some quantum constraints, the joint probabilities are not given, they are variables.
 Therefore, when writing down quantum constraints \eq{sdpc1}, we only need to consider the elements with the specific
value ($0$ and $1$) and the relation between different elements such
as some elements are the same. For convenience, instead of
$A_{\vec{x}}$ and $B_{\vec{y}}$, we use $a:a\in\tilde{A}$ and
$b:b\in\tilde{B}$ to denote Alice's and Bob's outcomes and $X(a)$
and $Y(b)$ are the associated measurement setting. The indexes $s,t$
of $\Gamma$ denote associated operators, i.e.,
$\Gamma_{a,b}=\Tr(E_{a}E_{b}\rho)$.

 For $\Gamma^{(1)}$, the associated quantum constraints are
\begin{itemize}
  \item $\Gamma^{(1)}_{1,1}=\Tr(\rho)=1$.
  \item  $\Gamma^{(1)}_{a,a'}=\delta_{aa'}\Gamma^{(1)}_{1,a}$ if
  $X(a)=X(a')$.
  \item $\Gamma^{(1)}_{b,b'}=\delta_{bb'}\Gamma^{(1)}_{1,b}$  if
  $Y(a)=Y(a')$.
  \item
  $\Gamma^{(1)}_{s,t}=\Gamma^{(1)}_{t,s}$.
\end{itemize}

We reexpress $\Gamma^{(1+AB)}$ by $4$ sub-matrixes,
\begin{eqnarray}
 \left(
  \begin{array}{cc}
    v_{1,1} & v_{1,2} \\
    v_{2,1} & v_{2,2} \\
  \end{array}
\right)
\end{eqnarray}
Since $\Gamma^{(1+AB)}$ is symmetric matrix, the sub-matrix
$v_{2,1}$ is equal to the transpose of $v_{1,2}$, and both
sub-matrix $v_{1,1}$ and $v_{2,2}$ are symmetric matrixes. Note
that, $v_{1,1}=\Gamma^{(1)}$. The elements of matrices $v_{1,2}$ and
$v_{2,2}$ are constrained by following quantum constrains:
\begin{itemize}
      \item
      $\Gamma^{(1+AB)}_{1,ab}=\Gamma^{(1+AB)}_{a,ab}=\Gamma^{(1+AB)}_{a,b}=\Gamma^{(1+AB)}_{b,ab}$.
      \item
      $\Gamma^{(1+AB)}_{ab,a'b}=\Gamma^{(1+AB)}_{a,a'b}=\Gamma^{(1+AB)}_{a',ab}$.
     \item
     $\Gamma^{(1+AB)}_{ab,ab'}=\Gamma^{(1+AB)}_{b,ab'}=\Gamma^{(1+AB)}_{b',ab}$.
 \item $\Gamma^{(1+AB)}_{a,a'}=0$, $\Gamma^{(1+AB)}_{a,a'b}=0$, and $\Gamma^{(1+AB)}_{ab,a'b'}=0$ if  $X(a')=X(a)$.
 \item $\Gamma^{(1+AB)}_{b,b'}=0$, $\Gamma^{(1+AB)}_{b,ab'}=0$, and $\Gamma^{(1+AB)}_{ab,a'b'}=0$ if
 $Y(b)=Y(b')$.
\item
  $\Gamma^{(1+AB)}_{s,t}=\Gamma^{(1+AB)}_{t,s}$.
\end{itemize}

\subsection{Estimating the number of constrains for $n=1$ and $n=1+AB$ certificates}

Due to the limitation of computer memory, we need to estimate the
number of these quantum constraints for different $k$ and $d$ RAC
protocols. The dimension of $\Gamma^{(1)}$ is $1+(d-1)(d^{k-1}+k)$,
we denote it as $dim$. The number of conditions corresponding to
different quantum behaviors is as follows.
\begin{center}
\begin{tabular}{|c|c|c|c|c|}
\hline
n=1                & symmetric matrix& $\Tr(\rho)=\Gamma^{(1)}_{1,1}=1$         & orthogonality          & $E_{a}E_{a}=E_{a},E_{b}E_{b}=E_{b}$          \\
\hline
number & $\frac{dim(dim-1)}{2}$ & $1$                    &$\frac{(d-1)(d-2)}{2}(d^{k-1}+k)$& $dim-1$ \\
\hline
\end{tabular}
\end{center}

The dimension of $\Gamma^{(1+AB)}$ is
$1+(d-1)(d^{k-1}+k)+(d-1)(d^{k-1}k)$, we denote it as $dim_{1+AB}$.
The number of conditions corresponding to different quantum
behaviors is as follows.
\begin{center}
\begin{tabular}{|c|c|c|c|c|c|}
\hline
n=1+AB                 & symmetric matrix& $Tr(\rho)=\Gamma^{1}_{1,1}=1$         & orthogonality          & $E_{a}E_{a}=E_{a},E_{b}E_{b}=E_{b}$  & same        \\
\hline
number & $\frac{dim_{1+AB}(dim_{1+AB}-1)}{2}$& 1            &$otha+othb+othc$& $dim_{1+AB}-1$ &$\sum_{i=1}^{7}same_{i}$\\
\hline
\end{tabular}
\end{center}

The quantum constraints orthogonality and commutativity make some
elements of certificate to be $0$ or to be the same. We will specify
to estimate the number of these special elements in $n=1+AB$
certificate. First, we estimate the number of elements whose value
is zero.
\begin{itemize}
\item The variable $otha=\frac{(d-1)(d-2)}{2}(d^{k-1}+k)$ is used to specify the number of zero elements for right upper matrix of $v_{1,1}$.
\item The variable $othb=2(d-1)^{2}(d-2)kd^{k-1}$ is used to specify the number of zero elements for sub-matrix $v_{1,2}$.
\item The variable $othc=\frac{kd^{k-1}(d-1)^{2}}{2}((d-2)(d-1)(d^{k-1}+k-2)+(d-1)^{2}-1)$ is used to specify the number of zero elements for right upper matrix
of $v_{2,2}$.
\end{itemize}

We estimate the variable $same_{i}$ which is used to denote the
number of equal pairs.
\begin{itemize}
      \item $\Gamma^{(1+AB)}_{1,ab}=\Gamma^{(1+AB)}_{a,ab}$,
$same_{1}=(d-1)^{2}(d^{k-1}k)$.
      \item $\Gamma^{(1+AB)}_{a,ab}=\Gamma^{(1+AB)}_{a,b}$,
$same_{2}=(d-1)^{2}(d^{k-1}k)$.
      \item $\Gamma^{(1+AB)}_{b,ab}=\Gamma^{(1+AB)}_{a,b}$,
$same_{3}=(d-1)^{2}(d^{k-1}k)$.
      \item $\Gamma^{(1+AB)}_{ab,a'b}=\Gamma^{(1+AB)}_{a,a'b}$,
$same_{4}=(d-1)^{3}d^{k-1}k(d^{k-1}-1)/2$.
     \item $\Gamma^{(1+AB)}_{a,a'b}=\Gamma^{(1+AB)}_{a',ab}$,
$same_{5}=(d-1)^{3}d^{k-1}k(d^{k-1}-1)$.
     \item $\Gamma^{(1+AB)}_{ab,ab'}=\Gamma^{(1+AB)}_{b,ab'}$,
$same_{6}=(d-1)^{3}d^{k-1}k(k-1)/2$.
    \item $\Gamma^{(1+AB)}_{b,ab'}=\Gamma^{(1+AB)}_{b',ab}$,
$same_{7}=(d-1)^{3}d^{k-1}k(k-1)$.
\end{itemize}

After estimating the number of conditions, we can think how to write
down these conditions with minimal computer memory. Here, we use the
numerical package named CVXOPT \cite{CVXOPT} to calculate the bounds
of the Bell-type inequalities. The primal problem of the cone
programming defined in CVXOPT is
\begin{subequations}\label{cvxopt}
\begin{align}
minimize \qquad &  c\cdot x\\
subject \quad to \qquad & Ax-b=0\\
                        & h-Gx\geq 0
\end{align}
\end{subequations}
Given $c$, $h$ which are the vectors and $A$, $G$ which are
matrixes, we can optimize the linear combination $c\cdot x$. Here
matrix $G$ is used to specify the positive definiteness constraint.
Writing down the positive definiteness constraint of a matrix $Z$
whose size is $s\times s$, we need the matrix $G$ with size
$s^{2}\times n$ to define the condition (where $n$ is the number of
variables $x$). That means, if we reduce the number of variables, we
can save the computer memory. To do this, we define the same
variable for two elements instead of constraining two variables with
the same value. On the other hand, if the value of some elements are
zero, it could also reduce the number of variables.

After using the conditions to reduce the number of variables, we can
estimate the number of variables in the certificate.\\
The number of variables in $\Gamma^{(1)}$ for different RAC
protocols:
\begin{center}
\begin{tabular}{|c|c|c|c|c|}
\hline
n=1                & d=2         & d=3          & d=4  & d=5        \\
\hline
k=2                & 10           &   50           & 153  &   364        \\
\hline
k=3                &28           &    288          & 1596      &   6160        \\
\hline
k=4                & 78           &   1922           &    20706   &  132612         \\
\hline
\end{tabular}
\end{center}
The number of variables in $\Gamma^{(1+AB)}$ for different RAC
protocols:
\begin{center}
\begin{tabular}{|c|c|c|c|c|}
\hline
n=1+AB                & d=2         & d=3          & d=4  & d=5        \\
\hline
k=2                & 15           &  182            &  1287   &  5964        \\
\hline
k=3          & 82           &     4068       & 61860      &474160      \\
\hline
k=4                & 486           &    71258          &1995810     &   24012612        \\
\hline
\end{tabular}
\end{center}

Due to the constraint of the computer memory (128GB), we could not
find the bounds of the Bell-type inequalities for arbitrary RAC
communication protocols. We find the bound what we can do and show
the result in the main text.

\end{appendix}


 \end{document}